\documentstyle[12pt,aps,epsf]{revtex}

\def\be{\begin{equation}}
\def\ee{\end{equation}}

\newcounter{fig}

\catcode`\@=11   

\newcommand{\fcaption}[1]{\vspace{1ex}   
        \refstepcounter{figure}   
        \setbox\@tempboxa = \hbox{\footnotesize {\bf Fig.~\thefigure.} #1}   
        \ifdim \wd\@tempboxa > 12cm   
           {\begin{center}   
        \parbox{12cm}{\footnotesize\baselineskip=8pt {\bf Fig.~\thefigure.} #1}   
            \end{center}}   
        \else   
             {\begin{center}   
             {\footnotesize {\bf Fig.~\thefigure.} #1}   
              \end{center}}   
        \fi}

\begin{document}

\title{Kinetics of stochastically-gated 
diffusion-limited reactions and geometry
of random walk trajectories
}

\vspace{0.9in}

\author{O.B{\'e}nichou, M.Moreau and  G.Oshanin\footnote{To whom the correspondence
should be addressed.} 
}

\vspace{0.7in}

\address{Laboratoire de Physique Th{\'e}orique des Liquides (CNRS - UMR 7600), \\
Universit{\'e} Pierre et Marie Curie,\\
4 place Jussieu, 75252 Paris Cedex 05, France
}

\begin{tightenlines}
                 
\vspace{0.9in}

\address{\rm (Received: )}
\address{\mbox{ }}
\address{\parbox{18cm}{\rm \mbox{ }\mbox{ }
In this paper we study the 
kinetics of diffusion-limited, pseudo-first-order
$A + B \to B$ reactions in 
situations 
in which the particles' intrinsic
reactivities
are not constant but vary randomly in time.
That is, 
we suppose that the particles 
are bearing "gates" which fluctuate in time, 
 randomly and independently of each other,
between two states - 
an active state, when the reaction may take place between  
an $A$ and a $B$ particles appearing at a close contact,
and a blocked state, when the reaction is completly inhibited. 
We focus here on two customary limiting cases of pseudo-first-order reactions -  
the so-called target annihilation and the Rosenstock trapping model, and
consider four different particular models, such that  
the $A$ particle can be either mobile or immobile, gated or ungated,
as well as ungated or gated $B$ particles can be fixed 
at random positions or move randomly.   
All models are formulated on a $d$-dimensional regular 
lattice and we suppose that the mobile species perform
independent, homogeneous, discrete-time lattice random walks. 
The model involving a single, immobile, ungated target $A$ and a concentration
 of mobile,
gated $B$ particles is solved exactly. For the remaining three models 
we determine exactly, in form of rigorous
lower and upper bounds showing the same $N$-dependence, 
the large-$N$
asymptotical behavior of 
the probability that the $A$ particle survives 
until the $N$-th step. We also realize 
that for all four models studied here
the 
$A$ particle survival probalibity 
can be interpreted
 as the moment generating function of some functionals of random walk
trajectories, such as, e.g., the number of self-intersections, the number of sites
visited exactly a given number of times, "residence time" on a random array of lattice
sites and etc. Our results thus apply to the asymptotical behavior
of the corresponding generating functions which has not been known as yet. 
}}
\address{\mbox{ }}
\address{\parbox{14cm}{\rm PACS No:  05.40+j, 05.60.+w, 02.50.Ey, 82.20.-w }}
\maketitle

\makeatletter
\global\@specialpagefalse

\makeatother

\end{tightenlines}

\pagebreak

\section{Introduction}

Many of naturally occuring chemical reactions, or reactions 
used in various
technological and material processing 
operations involve molecules with a rather complex 
internal structure.  
For such reactions the geometrical complexity of the molecules appears to be a significant rate-controlling factor,
additionally to the transport limitations 
and the constraints imposed by the
elementary reaction act;  that is, chemically   
active groups of complex molecules involved may be effectively screened by
its inactive parts, 
which thus impedes the access of the reactive species and inhibits the reaction. 
For instance, 
geometrical restrictions are crucial for ligand 
binding to proteins, such as, e.g. 
myoglobin or hemoglobin \cite{perutz}. Here, in the static $X$-ray
 structure of myoglobin there is no hole for
the ligand to enter and it is believed 
that binding of the 
ligand happens when the side chains blocking the
entrance swing out in course of their thermal motion (Figs.1 and 2). 
Similarly, 
intercalation of drugs by DNA may be controlled by
 breathing motions that involve the unstacking
of adjacent pairs of bases. In some cases, ligands themselves 
can possess a complicated
internal structure, as exemplified, for instance, by the
peptides, such that their reactivity may be influenced 
significantly by conformational changes. 
Last but not least, geometrical restrictions do 
manifest themselves 
in such 
contexts, as, e.g., 
certain chemical reactions occuring within
 biomembranes \cite{caceres}, incoherent exciton trapping 
 by substitutional traps on aromatic vinyl polymers \cite{igor}, molecular transport inside proteins
\cite{agmon,zwanzig,wang,klafter}, 
as well as in some medical therapies  \cite{spouge}. Clearly, 
understanding of the impact of
the geometrical limitations on the reaction kinetics constitutes an important
 challenge for the theoretical analysis.

A physically plausible
approach to account for the influence of the 
geometrical restrictions on the reactivity of complex molecules 
is to assume
that the reaction in question, say, a generic pseudo-first-order reaction
of the form
\be 
\label{reaction}
A + B \to B,
\ee
is modulated by the side reactions of the form $A \leftrightarrow A^* $ or 
$B  \leftrightarrow B^*$, 
where $A$ ($B$) stands for
an active, while $A^*$ ($B^*$) - for an inactive, blocked state, 
in which case the reaction in Eq.(\ref{reaction}) is complete inhibited. 
In other words, one says that one or both species involved in the reaction
are being gated, the gates changing in time 
their states according to some prescribed
rules.

Kinetics of gated diffusion-limited reactions has been studied
 analytically for nearly two decades.
 Following the seminal work of McCammon and Northrup \cite{mccammon}, 
who have analysed a simple
case of a two-state gating described by arbitrary deterministic function of time, 
several important advancements have been made. 
In particular,  classical Smoluchowski approach \cite{smol}  has
been generalized in Refs.\cite{szabo1,szabo2,ber1} 
to describe kinetics of $\em stochastically$-gated (SG) pseudo-first-order
reactions in Eq.(\ref{reaction}) for both cases when the gate is imposed
 on the traps $B$ or
on the $A$ particles. It has been realized that no symmetry exists between these
 two situations; as a matter of fact,
the kinetic behavior appears to be rather different depending on which of the species precisely 
is being gated. Furthermore, 
Spouge et al \cite{spouge2} studied the 
SG reactions with more general mechanisms, 
including a non-Markovian case,
while Berezhkovski with collaborators
discussed the impact of the many-particle effects 
on  the  SG reactions kinetics \cite{ber1,ber2,ber3}.
However,  the available theoretical analysis 
is either based on uncontrollable assumptions, 
akin to the Smoluchowski-type  approaches (see, e.g. \cite{weiss,gleb} and references
therein), or
employs exact formalisms, which  do not always allow for explicit calculation of
 the corresponding decay patterns and become tractable only when some simplifying assumptions are made.
Consequently, except for the relatively simple model involving an immobile
ungated target $A$ and a concentration of mobile SG $B$ particles (see, e.g.
Refs.\cite{szabo1,szabo2,ber1,spouge2}), 
temporal evolution of the SG  reactions 
remains incompletely understood. 
This is, of course, not at all surprising since diffusion-limited reactions with
stochastic reactivity 
clearly pose more complex technical problems to the theoretical analysis
 than their ungated
counterparts, which themselves are not solvable exactly and often exhibit spectacular,
essentially many-particle behavior.  

In this paper we study in detail the
kinetics of the SG pseudo-first-order
reactions in Eq.(\ref{reaction}),
involving a single $A$ particle
and a concentration of  $B$ particles.
We focus here on two limiting cases - 
the so-called target annihilation
\cite{blumen1,blumen,burlatsky,szabo3,redner} and the Rosenstock trapping model
\cite{rosenstock}, and
consider four different models such that
an $A$ particle can be either mobile or immobile, gated or ungated,
as well as ungated or gated $B$ particles can be fixed at random positions or move randomly  (Fig.3).   
For computational convenience, 
all models are formulated on a $d$-dimensional regular lattice and we suppose that mobile species perform
independent, homogeneous, discrete-time lattice random walks. 
Further on, in regard to the reactivity fluctuations 
we restrict ourselves to the
two-state Poisson gating model of Ref.\cite{spouge2}, 
in which  each gate is supposed to be in either of two states - an active and a blocked one, and to
update its state at each tick of the clock. The updating process is assumed 
to proceed completely at random, without memory in time and 
without 
correlations with the gates imposed on the other particles. 

For such a gating model we analyse the time
 evolution of the  $A$ particle
survival probability $P_N$,  i.e. the probability
 that a single mobile or immobile, gated or ungated $A$ 
particle is not annihilated
up to the $N$-th time step by a concentration of immobile or mobile, 
ungated or gated $B$ particles. For the model involving a single immobile, ungated
target $A$ and a concentration of SG mobile $B$ particles the complete temporal
evolution of $P_N$ is calculated exactly. For the remaining three models we  
determine exactly the form of the large-$N$ asymptotical decay
 of $P_N$ by deriving rigorous
lower and upper bounds, which both 
show the same $N$-dependence but slightly differ in prefactors.

An interesting by-product of our analysis, which appears in especially lucid fashion within the
discrete-space description, is that for any particular model
 the $A$ particle survival probability $P_N$ can be interpreted as a moment  generating function of certain
functionals, 
which mirror the internal geometry of random walks trajectories - an issue which has recently 
gained a renewed
attention in view of many important applications \cite{hughes,duplantier,lawler,nechaev}.
As a matter of fact, it has been known
for a long time 
that for ungated trapping reactions taking place on a $d$-dimensional 
lattice the survival probability $P_N$ can be thought off as
 the moment generating function of the
number ${\cal S}\Big(\{{\bf r}_N\}\Big)$ of distinct sites  visited by a given trajectory $\{{\bf r}_N\}$ of an 
$N$-step lattice random walk \cite{rosenstock}; as well,  
for Brownian motion in a $d$-dimensional continuum $P_N$ can be interpreted
as the generating function
of the volume of the so-called Wiener sausage, 
i.e. the volume swept by a spherical particle
during time $t$ \cite{donsker}.  We realize that 
for the SG pseudo-first-order reactions
some other characteristic 
functionals of random walk trajectories come into play, which probe 
some interesting aspects of the geometrical 
structure of a single or of a collection of
lattice random walks.  Depending on the particular 
model $P_N$ appears then as the generating function of 
such characteristic functionals of random walk trajectories as, e.g.,  the number of self-intersections, the number
of sites visited exactly a given number of times, "residence time" on a random array 
of lattice sites, as well as some
others.   
Consequently, our results also apply to the 
asymptotical behavior of the corresponding generating functions, which 
has not been known as yet in many cases.

The paper is structured as follows. In section II we discuss two customary limiting cases
 of the psedo-first-order
reactions in Eq.(1) - the so-called target annihilation model and the Rosenstock trapping model, and 
present a brief outline of different results concerning their kinetic
behavior in the ungated case. In section III
 we consider a gated target problem,
focusing on the survival probability $P_N$ 
of an immobile ungated target in the presence of randomly moving, stochastically-gated
$B$ particles. We derive here an exact temporal evolution 
of the probability that none of mobile particles hits the target
within an $N$-step walk. Further on, in section IV we study the survival of an immobile
 $\em gated$ target in the presence
of a concentration of mobile particles. 
The corresponding survival probability is found here in form of rigorous lower and upper 
bounds,
which both display the same $N$-dependence and thus determine the temporal evolution of $P_N$ exactly. 
Next, in sections V and VI
we consider two models of stochastically-gated trapping reactions; 
in the first case these are the immobile 
traps $B$ who are supposed to be stochastically
gated, while the mobile $A$ particle is assumed to be always in the reactive state (ungated), 
and in the second case the mobile $A$
particle is assumed to bear
 a stochastic gate, while the immobile traps $B$ are being considered as perfect, non-fluctuating
traps. For both cases the $A$ particle survival probability is determined exactly, in form of rigorous lower and
upper bounds showing the same $N$-dependence. Finally, we conclude in section VII
 with a brief summary of our results.

\section{A reminder on the kinetics of ungated pseudo-first-order reactions: target annihilation and the
Rosenstock trapping models.
}

To fix the ideas, we present first
a brief summary of results concerning the kinetics of
 ungated, diffusion-limited pseudo-first-order reactions in Eq.(\ref{reaction})
  taking place on
$d$-dimensional regular lattices.  We will focus here 
and in the remainder of the paper on
two particular cases - the so-called target annihilation model, involving a single,  immobile $A$
particle and a concentration of randomly moving $B$ particles, and  the Rosenstock trapping model
involving a single mobile $A$
particle performing lattice random walk in the presence of a concentration of
immobile, randomly placed $B$
particles - the traps. In subsequent sections of our work we will study how the kinetics in these two
models will be modified if one imposes a stochastic gate on either of two species; consequently, the
results of this section will serve us in what follows as an important point of reference.

\subsection {Target  annihilation.} 

We start with the target annihilation model
 which allows for an exact solution \cite{blumen1,blumen,burlatsky,szabo3}. 
Consider an immobile $A$ particle, located at the lattice origin, 
and $B$ particles,  which are initially all placed at random positions 
on a $d$-dimensional regular
lattice and after that are allowed to move by 
performing independent, homogeneous, discrete-time random walks.
 As soon as any of the $B$s appears at
the lattice origin, the $A$ particle gets instantaneously 
annihilated. Thus, following
the earlier introduced terminology, the $A$ particle will be called here as the
"target", while the $B$ particles will be referred to as 
the "scavengers" \cite{redner}.
The property whose temporal evolution we wish to study is the
probability $P_N = \exp[ - c Q_N^{(tar)}]$ that 
the target particle "survives" until time $N$; here and henceforth $c$ will denote 
the number density of the $B$ particles, while $Q_N^{(tar)}$, in view of 
the pseudo-first order of the reaction in Eq.(\ref{reaction}),
can be interpreted as the 
integral effective reaction rate.

Now we define the model more precisely. We first 
suppose that the lattice is of a finite extent and contains $M$
sites, whereas
the  number of the scavengers is also fixed
and is equal to $K$.
In what follows we will turn to the limit $M,K \to \infty$, while keeping the
ratio $K/M$ fixed,  $K/M = c$.

Further on, let ${\bf S}_n^{(k)}$  denote the position
 at which the $k$-th scavenger appears
on the $n$-th step ($n = 0,1, \ldots, N$) for a given realization of its trajectory. 
Then, we construct a function $\Psi_N$ of the form
\begin{eqnarray}
\Psi_N&=&\lim_{\beta\to\infty}\exp\left[-\beta \; \sum_{n=0}^N\sum_{k=1}^K{\cal
I}\left({\bf S}_n^{(k)}\right) \right] = \nonumber\\
&=&\prod_{n=0}^N\prod_{k=1}^K\lim_{\beta\to\infty}\exp\left[-\beta \; {\cal
I}\left({\bf S}_n^{(k)}\right) \right],
\label{eq:1}
\end{eqnarray}
where ${\cal
I}\left({\bf S}_n^{(k)}\right)$  is the indicator function, 
\begin{equation}
\label{indic}
{\cal
I}\Big({\bf X}\Big) =  \left\{\begin{array}{ll}
1,  \;    \mbox{ ${\bf X} = 0,$}\\
0, \;     \mbox{${\bf X} \neq 0,$}
\end{array}
\right.
\end{equation} 
which shows whether the $k$-th scavenger 
is at the lattice origin $0$ at the $n$-th step, or elsewhere. 

Note, that
the function $\Psi_N$ serves as the indicator of the reaction event;  
it is equal to one if within an $N$-step walk none of the $K$ scavengers
 has visited the origin,
 (i.e. the target),
 and turns to zero when 
within the "time" interval $[0,N]$ at least one of the scavengers has at least once
visited the origin. 

The property one is generally interested to compute 
is not, however, the realization-dependent function 
$\Psi_N$, but rather its averaged value $P_N = <\Psi_N>$, the average being taken
 over different realizations of scavengers' random
walks and their initial positions. 
Below we will briefly outline an exact computation of  $<\Psi_N>$ 
(see also Refs.\cite{blumen1,blumen,burlatsky,szabo3,redner}).

Noticing first that all ${\bf S}_n^{(k)}$ with different values of 
$k$ are independent of each other, we have 
that the target survival probabilitity can be written in the following factorized form
\be
\label{rul}
P_N = \Big({\rm lim}_{\beta \to \infty} \frac{1}{M} 
\sum_{{\bf S}_0} {\rm E}_{{\bf S}_0}\Big\{\exp\Big[ - \beta \; \sum_{n = 0}^N 
{\cal
I}\left({\bf S}_n\right)\Big]\Big\}\Big)^{K},
\ee
where the summation extends over all sites of a $d$-dimensional lattice, while 
the symbol ${\rm E}_{{\bf S}_0}\Big\{ \ldots \Big\}$ denotes the expectation 
on the set of different random walk trajectories starting at the site ${\bf S}_0$. Note next that 
\be
\label{ru}
{\rm E}_{{\bf S}_0}\Big\{{\rm lim}_{\beta \to \infty} \exp\Big[ - \beta \; \sum_{n = 0}^N 
{\cal
I}\left({\bf S}_n\right)\Big]\Big\} \equiv {\rm Prob}\Big({\bf S}_n \neq 0 \; \text{for any } 
n \in [0,N]| {\bf S}_0\Big),
\ee
i.e. is equal to the probability that a random walker starting at the site ${\bf S}_0$ does not visit the 
origin
within first $N$ steps.
Turning now to the infinite-space limit, i.e. letting $K,M \to \infty$ while keeping their ratio fixed, $K/M =
c$, we find then that  Eq.(\ref{rul})  attains the
form
\begin{eqnarray}
\label{target}
P_N &=& lim_{K,M \to \infty| K/M = c}\Big(1 - \frac{1}{M} \sum_{{\bf S}_0} \Big[1 - {\rm Prob}\Big({\bf S}_n \neq 0 \; 
\text{for any } n \in [0,N]| {\bf S}_0\Big)\Big]\Big)^K = \nonumber\\
&=& \exp\Big[ - c \sum_{{\bf S}_0} {\rm Prob}_N\Big(0| {\bf S}_0\Big) \Big],
\end{eqnarray}
where $ {\rm Prob}_N\Big(0| {\bf S}_0\Big)$ stands for  
the probability  that a first passage from the site  ${\bf S}_0$ to the origin did
actually occur within the first $N$ steps. 

The probability ${\rm Prob}_N\Big(0| {\bf S}_0\Big)$ is known exactly (see, e.g.
\cite{hughes});  
being  
summed 
over all possible starting points it defines another important characteristic property of
 random walks -   the expectation of the number ${\cal S}\Big(\{{\bf S}_N\}\Big)$ of $\em distinct$
 sites visited by a given trajectory $\{{\bf S}_N\}$ 
of an $N$-step walk starting at the origin, i.e., 
\be
\label{lu}
\sum_{{\bf S}_0} {\rm Prob}_N\Big(0| {\bf S}_0\Big) =  {\rm E}_{0}\Big\{{\cal S}\Big(\{{\bf S}_N\}\Big)\Big\}
\ee
The expected number of distinct sites visited by an $N$-step walk shows different
asymptotical behavior depending on the dimensionality $d$ and on the type 
of the lattice  \cite{hughes}. In particular, for $d$-dimensional 
Polya random walks one has 
\begin{eqnarray}
\label{visits}
&&d = 1 \; \; \;   {\rm E}_{0}\Big\{{\cal S}\Big(\{{\bf S}_N\}\Big)\Big\} = \Big(\frac{8 N}{\pi}\Big)^{1/2} + 
{\mathcal O}\Big(\frac{1}{\sqrt{N}}\Big), \nonumber\\
&&d = 2 \; \; \;   {\rm E}_{0}\Big\{{\cal S}\Big(\{{\bf S}_N\}\Big)\Big\} = \frac{\pi \; C_2 \; N}{\ln(N)}
 + 
{\mathcal O}\Big(\frac{N}{\ln^2(N)}\Big), \nonumber\\
&&d = 3 \; \; \;  {\rm E}_{0}\Big\{{\cal S}\Big(\{{\bf S}_N\}\Big)\Big\} = \frac{N}{P(0|0; 1^{-})} + 
{\mathcal O}\Big(\sqrt{N}\Big), 
\end{eqnarray}
where $C_2 = 4/3 \sqrt{3}$, $1$ and $2/\sqrt{3}$ for hexagonal, square and triangular two-dimensional 
lattices, respectively.  The constant $P(0|0; 1^{-})$ determines the probability $R$ 
of eventual return to the origin, $P(0|0; 1^{-}) = \Big(1 - R\Big)^{-1}$, and is defined
as the limit $\xi \to 1^-$
of the generating
function  $P(0|0;\xi) = 
 \sum_{N=0}^{\infty}P_N(0|0) \xi^{N}$, $P_N(0|0)$ being the probability  of 
finding a random walker at the origin at the $N$-th step, provided that the walk has started 
at the origin. The exact values of $P(0|0; 1^{-})$ are 
$12 \Gamma^6(\frac{1}{3})/(2^{4/3} \pi^4)$, $\sqrt{6} \; \Gamma(\frac{1}{24}) \;
\Gamma(\frac{5}{24}) \; \Gamma(\frac{7}{24}) \; \Gamma(\frac{11}{24})/(284 \; \pi^3)$, 
$\Gamma^4(\frac{1}{4})/(4 \; \pi^3)$
and $\Gamma^6(\frac{1}{3})/(2^{14/3} \pi^4)$ for diamond, simple cubic, body-centered cubic and face-centered cubic
lattices, respectively \cite{hughes}. 

Therefore, for the target annihilation model the decay law can be computed exactly and 
the integral effective reaction rate $Q_N^{(tar)}$ equals simply
 the expected number of distinct sites visited by an $N$-step walk \cite{blumen1,blumen,burlatsky,szabo3}.
 Note also that the decay form 
appears to be essentially dependent on the dimensionality of the embedding lattice; it is characterized by a
stretched-exponential dependence for low dimensional lattices, on which the Polya walks are recurrent ($R = 1, 
P(0|0; 1^{-}) = \infty$), and
shows a purely exponential behavior for lattices of spatial dimension $d > 2$, where $R < 1$ and
$P(0|0; 1^{-})$ is well defined.

\subsection{Rosenstock trapping model.} 

We turn next to the so-called Rosenstock trapping 
problem, in which one focuses on the fate of a single
$A$ particle performing a random walk over the lattice in the presence of immobile, perfect, randomly placed 
traps $B$. 

We start assuming again that the lattice is finite and contains $M$ sites. The $K$ traps $B$ are placed completely
at random and their positions are determined by the lattice-vectors ${\bf S}^{(k)}$, $k = 1,2, \ldots ,K$.
Denoting the lattice position of the $A$ particle at the 
 $n$-th step as ${\bf r}_n$, we can now write down
the indicator function of
the reaction event as  follows
\be
\label{function}
\Psi_N = {\rm lim}_{\beta \to \infty} \exp\Big[- \beta \; \sum_{n=0}^{N} \sum_{k=1}^{K}
 {\cal I}\Big({\bf r}_n - {\bf S}^{(k)}\Big)\Big],
\ee  
where ${\cal I}\Big({\bf X}\Big)$ is the indicator function defined in Eq.(\ref{indic}).  The function in 
Eq.(\ref{function}) is equal to one for such $N$-step trajectories which avoid passing through any of the sites 
 ${\bf S}^{(k)}$ and turns to zero for those trajectories which visit at least once at least one of these sites.

Now, we pass to averaging the function in Eq.(\ref{function}) over the traps' placement.
Since all  ${\bf S}^{(k)}$ are
mutually-independent, one can wright down such an the average in the factorized form
\begin{eqnarray}
<\Psi_N> &=& {\rm E}_{0}\Big\{\prod_{k = 1}^K \Big( \frac{1}{M} \sum_{{\bf S}^{(k)}}
 {\rm lim}_{\beta \to \infty} 
\exp\Big[- \beta \; \sum_{n=0}^{N} 
 {\cal I}\Big({\bf r}_n - {\bf S}^{(k)}\Big)\Big] \Big) \Big\}
\end{eqnarray} 
Next, 
in the limit  $K,M \to \infty$ one has that
\begin{eqnarray}
<\Psi_N> &=& {\rm E}_{0}\Big\{ \Big( \frac{1}{M} \sum_{{\bf S}} {\rm lim}_{\beta \to \infty} 
\exp\Big[- \beta \; \sum_{n=0}^{N} 
 {\cal I}\Big({\bf r}_n - {\bf S}\Big)\Big]\Big)^K\Big\} = \nonumber\\
&=& {\rm E}_{0}\Big\{\exp\Big[- c \sum_{{\bf S}} \Big(1 - {\rm lim}_{\beta \to \infty}
\exp\Big[ - \beta \; \sum_{n=0}^{N} 
 {\cal I}\Big({\bf r}_n - {\bf S}\Big)\Big]\Big)\Big]\Big\}= \nonumber\\
&=& {\rm E}_{0}\Big\{\exp\Big[- c \sum_{{\bf S}} \Big(1 - 
{\cal I}\Big(\sum_{n=0}^{N}{\cal I}\Big({\bf r}_n - {\bf S}\Big)\Big)\Big)\Big]\Big\},
\end{eqnarray} 
where
\be
\sum_{{\bf S}} \Big(1 - 
{\cal I}\Big(\sum_{n=0}^{N}{\cal I}\Big({\bf r}_n - {\bf S}\Big)\Big)\Big) \equiv 
{\cal S}\Big(\{{\bf r}_N\}\Big),
\ee
is the number of distinct sites visited by a given trajectory $\{{\bf r}_N\}$. Consequently, 
the $A$ particle survival probability  obeys
\be
\label{sur}
<\Psi_N> = {\rm E}_{0}\Big\{\exp\Big[- c \; {\cal S}\Big(\{{\bf r}_N\}\Big)\Big]\Big\} = \exp\Big[ - c \; Q_N^{(tr)}\Big]
\ee
and hence, appears here as 
the moment generating function of the number of distinct sites visited by an $N$-step random
walk. 

Therefore, the major difference between 
the target annihilation model 
and the trapping model is exactly
that in the former the survival probability is the exponential of the
expected  number of distinct sites, while in the latter case it involves a fairly 
more complex property - its moment generating function.
In consequence, the trapping problem turns out to be 
 essentially more difficult than the target one and hence, shows a richer behavior.

To display the time evolution of $P_N$ in the trapping problem, we will first outline 
the predictions of a certain heuristic approach - the so-called Rosenstock approximation
\cite{rosenstock}, 
and then write down the results of a rigorous analysis by Donsker and Varadhan,
which concerns the large-$N$ asymptotical behavior \cite{donsker}.

\subsubsection{Rosenstock approximation.}

This approximation has been first applied
 by Rosenstock \cite{rosenstock} in his studies of luminescence
 quenching kinetics and amounts, in essence, to the mere replacement of the average of an exponential of 
the number of sites visited by an exponential of
the expected number,  
\be
\label{roro} 
<\Psi_N> \approx \exp\Big[- c \; {\rm E}_{0}\Big\{ {\cal S}\Big(\{{\bf r}_N\}\Big)\Big\}\Big]
\ee
 As one may readily notice, 
this heuristic procedure 
yields exactly the behavior found for the target annihilation problem and
consequently, within the framework of this approximation one finds  that 
$Q_N^{(tr)} \equiv Q_N^{(tar)}$.

As a matter of fact, 
numerical simulations demonstrate that a rather crude and uncontrollable 
approximation in Eq.(\ref{roro}) provides quite a fair
description of the decay for the trapping problem for intermediate 
values of $N$ \cite{blumen}, 
until at very large $N$ some deviations emerge.
The
reason why the Rosenstock approximation works at intermediate $N$ 
can be apparently explained as follows: As a matter of fact,  the Rosenstock approximation represents a rigorous lower bound on $P_N$, since replacement of
the average of an exponential of the number of  distinct sites visited by an exponential of
the expected number is tantamount to the application of the Jensen inequality, 
$<\exp[- c \;  Q_N]> \geq \exp[- c <Q_N>]$. On the other hand, this
inequality can be rewritten as
\be
< \Psi_N > = < \exp\Big[\ln(\Psi_N)\Big] > \geq \exp\Big[< \ln(\Psi_N) >\Big],
\ee
which signifies that in such an approach
 the average of the idicator function is approximated by the exponential
of the averaged logarithm of this function. Since logarithm
 is a slowly varying function, it is generally believed that 
its average value
is supported by typical realizations of random walk trajectories which are 
representative  
at moderate values of $N$.

\subsubsection{Fluctuation-induced large-$N$ tails of the survival probability.}

In the large-$N$ limit, however, kinetics of the trapping reactions proceeds somewhat slower 
than that
predicted by Eq.(\ref{roro}). This happens, namely,  due to some fluctuation effects, which a mean-field-type
 approximation
in Eq.(\ref{roro}) can not capture.  It has been first predicted 
in Ref.\cite{bal}, and
subsequently 
proven by Donsker and Varadhan \cite{donsker}, that for
arbitrary $d$ the decay follows
\be
\label{dv}
P_N \sim \exp[ - a_d c^{2/(d+2)} N^{d/(d+2)}], \; \; \; N \to \infty,
\ee
where $a_d$ is a constant, dependent on the lattice dimensionality \cite{donsker}.  

The physical origin of such an anomalous
decay law can be illustrated by the following heuristic derivation. 
Consider first the 
function in Eq.(\ref{sur}) 
and suppose that for some  given realization of the
$N$-step $A$ particle trajectory $\{{ \bf r}_N\}$ the maximal excursion from the origin is 
equal to $R_{max}$. Consequently, for this realization the number ${\cal S}\Big(\{{\bf r}_N\}\Big)$ 
of distinct sites visited by this realization 
of random walk trajectory  can be majorized as
${\cal S}\Big(\{{\bf r}_N\}\Big) \leq R_{max}^d$, and the overall decay function can be 
bounded from below by
\be
\label{esti1}
P_N \geq \int d^d  R_{max} \; \exp\Big[ - c  R_{max}^d\Big] \;
 {\rm Prob}(max \{{ \bf r}_N\} = R_{max}),
\ee
where ${\rm Prob}(max \{{ \bf r}_N\} = R_{max})$ is
 the probability that for a $d$-dimensional, $N$-step random
walk
the maximal displacement  from the starting point is exactly equal to 
$ R_{max}$. 
For sufficiently large $N$, the leading behavior of this probability follows 
$ \sim \exp\Big[ - \gamma_d N/R_{max}^2\Big]$, where $\gamma_d$ is a
dimension-dependent  constant. Substituting the latter form
 to Eq.(\ref{esti1}) and noticing that the integrand is a bell-shaped function, 
we thus perform the integration in terms of the saddle-point method. In doing so 
  one finds that 
the value of $R_{max}$ which provides the maximum to the integrand is given by $R_{max}^* = (2
\gamma_d N/c d)^{1/(d+2)}$. Consequently,  the overall decay function obeys
\be
\label{esti2}
P_N \geq  \exp\Big[ - 
const \;  
c^{2/(d+2)} \; N^{d/(d+2)} \Big],
\ee
which bound displays 
exactly the same $N$-dependence as the result in Eq.(\ref{dv}). 

Another illustrative derivation
can be performed starting directly from the 
definition of the indicator function of the reaction event $\Psi_N$ in Eq.(\ref{function}).  To do this, suppose first 
that for a given realization of
traps' placement the nearest to the origin trap $B$ appears at a certain distance $L$.  
For such a realization, evidently,
$\Psi_N = 1$ for those trajectories $\{{\bf r}_N\}$ which do not leave within $N$ first steps 
the volume $L^d$. Consequently, here the overall decay function can be bounded from below by
\be
\label{n}
P_N \geq max_L \Big\{ \exp\Big[- c L^d\Big] \;  {\rm Prob}(|{\bf r}_n| \leq L \; \text{for any} \; n \in
[0,N]|0) \Big\},
\ee
where the first multiplier gives the probability of having 
a trap-free void of volume $L^d$, while the second one stands
for the probability that a random walk starting at the origin does not leave this volume within first $N$
steps. For $N$ sufficiently large, the latter probability follows 
$\exp\Big[ - \gamma_d N/L^2\Big]$, 
and hence, maximizing the rhs of Eq.(\ref{n}) with respect to $L$, i.e. searching for the maximal lower bound,
one ends up with the dependence 
of essentially the same form as that given by Eq.(\ref{esti2}). 

Consequently, these bounds demonstrate that the long-time behavior of the survival
 probability in Eq.(\ref{dv})
is supported, first, by the presence of sufficiently large trap-free voids of typical 
size $\sim N^{1/(d+2)}$, and second, 
by realizations of random walks constrained not to leave these voids
 within time $N$, i.e. atypical,
spatially-confined realizations for which $|{\bf r}_N|$ 
grows in proportion to $N^{1/(d+2)}$ only.  

To conclude this section, we emphasize that the long-time behavior of the trapping and of the target annihilation problems
are different. Whereas the integral effective rate constant for the target annihilation follows the behavior of 
the expected number of distinct
sites visited, the trapping decay contains all the higher moments of this characteristic property
and tends at very long times towards the asymptotic form in Eq.(\ref{dv}), i.e.
$Q^{(tr)}_N \sim (N/c)^{d/(d+2)}$ as $N \to \infty$.
 Thus the decay patterns differ, depending on
which of the two species in Eq.(\ref{reaction})
is the mobile one. One has consequently a counter example to the view that in reaction kinetics only the relative motion of
the species, but not the individual movements are important.  In what follows we
intend to analyse how  the 
reaction kinetics will
differ  depending on which of the species precisely is being gated.

\section{Model I: an immobile target $A$ and randomly moving
 stochastically gated scavengers
$B$.}

We start our analysis of
 stochastically-gated pseudo-first-order reactions 
considering first survival of an immobile target  in the presence of a concentration of
mobile gated scavengers $B$ - the model which again admits an exact solution.
We suppose here that each gate can be in either of two states, one - active and the other one - blocked.
In the active state the $B$ particles are reactive, whereas the blocked or inactive
state inhibits the reaction. The gate on any $B$ particle is supposed to update its
 state at each moment
of time, at random, and independently of the gates imposed on the other particles. 
The $A$ particle gets "annihilated" at the very moment
when any of the $B$ particles visits 
it for the first time being in the reactive state. On contrary, if any of the $B$s visits the $A$ particle
being in the blocked state, both particles can harmlessly coexist with each other.

More precisely, we specify  the reactive ability of the $k$-th scavenger,
 ($k = 1, \ldots ,K$), at
the  $n$-th step, $n = 0,1,
\ldots, N$, by  assigning to each of the $B$ particles a 
random variable $\eta_{n}^{(k)}$. 
This random variable
 may assume two values - $0$ with probability $p$, in which case the scavenger
is  neutral
with respect to the reaction, 
and the value $1$ with probability $1 - p$, 
which corresponds to the reactive state. 
In all the models to be studied here, 
we will suppose that the
reactivities of gated particles follow independent Poisson processes \cite{spouge2}, 
such that   
all $\eta_{n}^{(k)}$  
are independent, delta-correlated with respect to
$n$ and $k$ randomly distributed variables. 
The average with respect to the distribution of $\eta_{n}^{(k)}$  
will be denoted
by the overbar.

Further on, denoting as ${\bf S}_n^{(k)}$ the position
 at which the $k$-th scavenger appears
at the $n$-th step
 for a given realization of its $N$-step trajectory $\{{\bf S}_N^{(k)}\}$, we 
construct the reaction event indicator function $\Psi_N$, which now takes the form
\begin{eqnarray}
\Psi_N&=&\lim_{\beta\to\infty}\exp\left[-\beta \; \sum_{n=0}^N\sum_{k=1}^K{\cal
I}\left({\bf S}_n^{(k)}\right)\eta_n^{(k)}\right] = \nonumber\\
&=&\prod_{n=0}^N\prod_{k=1}^K\lim_{\beta\to\infty}\exp\left[-\beta \; {\cal
I}\left({\bf S}_n^{(k)}\right) \; \eta_n^{(k)}\right],
\label{eq:1000}
\end{eqnarray}
where ${\cal
I}\Big(X\Big)$ is the indicator function
showing whether the $k$-th scavenger is at the lattice origin at the $n$-th step,  or elsewhere, Eq.(\ref{indic}). 

The indicator function $\Psi_N$ 
 is equal to one if within the $N$-step random walk 
none of  $K$ scavengers has visited the origin
being in the reactive state, and turns to zero otherwise, i.e. in case when 
within the interval $[0,N]$ at least one of the scavengers has once
visited the origin being in the
reactive state.  Its average over  the reactivity fluctuations, i.e. 
 the states of the gates $\eta_n^{(k)}$ can be
performed very directly, since for the Poisson gating model under study
all terms in the double product in Eq.(\ref{eq:1000}) are
statistically independent of each other. Consequently, we have that
\begin{eqnarray}
\overline{\Psi_N}&=&\prod_{n=0}^N\prod_{k=1}^K\overline{\lim_{\beta\to\infty}\exp\Big[-\beta \; {\cal
I}\left({\bf S}_n^{(k)}\right)\eta_n^{(k)}\Big]}\nonumber\\
&=&\prod_{n=0}^N\prod_{k=1}^K\left\{(1-p)\lim_{\beta\to\infty}\exp\Big[- \beta \; {\cal
I}\left({\bf S}_n^{(k)}\right) \Big]+p\right\}
\label{eq:2}
\end{eqnarray}
Further on, noticing that 
\begin{equation}
\lim_{\beta\to\infty}\exp\Big[- \beta \; {\cal
I}\left({\bf S}_n^{(k)}\right)\Big] = 1 - {\cal
I}\left({\bf S}_n^{(k)}\right)
\end{equation}
and hence, that
\begin{equation}
(1-p)\lim_{\beta\to\infty}\exp\Big[-\beta{\cal
I}\left({\bf S}_n^{(k)}\right)\Big]+p=\exp\Big[- \alpha_p \; {\cal
I}\left({\bf S}_n^{(k)}\right)\Big], \; \; \; \alpha_p = - ln(p),
\label{eq:4}
\end{equation}
we find that the  indicator function of the reaction event
$\Psi_N$,  averaged over the reactivity fluctuations, takes the form
\begin{equation}
\overline{\Psi_N}=\exp\Big[ - \alpha_p \; {\cal N}\left(\{{\bf S}_N^{(k)}\}\right)\Big],
\end{equation}
where
\be
{\cal N}\left(\{{\bf S}_N^{(k)}\}\right) = \sum_{n=0}^N\sum_{k=1}^K{\cal
I}\left({\bf S}_n^{(k)}\right) 
\label{eq:5}
\end{equation}
Note now that the functional ${\cal N}\left(\{{\bf S}_N^{(k)}\}\right)$ determines 
the number of times that $K$ given $N$-step random walk trajectories $\{{\bf S}_N^{(k)}\}$, with their starting points at
fixed positions ${\bf S}^{(k)}_0$,
pass through the origin. Note also that here
all walks contribute independently, which means that
the simultaneous visit of $k$ walkers is counted $k$ times, 
and hence, ${\cal N}\Big(\{{\bf S}_N^{(k)}\}\Big)$ can be interpreted as  
the "residence time" at the origin for 
$K$ independent random walkers.

As a matter of fact, the number ${\cal N}\left(\{{\bf S}_N^{(k)}\}\right)$
can be also viewed from a different perspective,
which turns to be rather useful 
for illustration of the distinction between different 
models. Namely, on Fig.4 we
depict, for a one-dimensional case, a 
given realization of several random walks trajectories in form of "directed polymers" in $1 +
1$-dimensions, the $Y$-axis being the time $n$; 
in this language ${\cal N}\left(\{{\bf S}_N^{(k)}\}\right)$ appears as
the total number of times that an $n$-axis is intersected on a
segment $[0,N]$ by a 
brush of 
$K$ phantom directed polymers with their ends fixed on a $d$-dimensional substrate. The survival
probability $P_N$ can be thought off 
 then as the generating function 
of the moments of the number of visits rendered by $K$
independent walkers to the origin during "time" $N$, or of 
the number of intersections of the $n$-axis on the
segment $[0,N]$ by a brush of directed polymers.  
In the infinite space limit $P_N$ is hence the generating function of
moments of the number of visits to the 
origin by random walkers which are initially uniformly distributed with a given
number density $c$ on an infinite
$d$-dimensional lattice.

We pass next to averaging over the scavengers' trajectories. 
Noticing that all multipliers in Eq.(\ref{eq:5}) with the same index $k$ are
again statistically independent of the multipliers with a different $k$, 
we may write down this average as
\begin{eqnarray}
P_N = <\overline{\Psi_N}>&=&<\prod_{k=1}^K  \exp\Big[  - \alpha_p \; \sum_{n=0}^N{\cal
I}\left({\bf S}_n^{(k)}\right)\Big]> = \nonumber\\
&=& \prod_{k=1}^K <\exp\Big[ - \alpha_p \; \sum_{n=0}^N{\cal
I}\left({\bf S}_n^{(k)}\right)\Big]> = \nonumber\\
&=&\left(\frac{1}{M}\sum_{{\bf S}_0}{\rm E}_{{\bf S}_0}\left\{\exp\Big[ - \alpha_p \; \sum_{n=0}^N{\cal
I}\left({\bf S}_n\right)\Big]\right\}\right)^K 
\label{eq:6}
\end{eqnarray}
Turning to the infinite space limit, we find from Eq.(\ref{eq:6}) 
that the survival probability $P_N$ follows
\begin{eqnarray}
P_N  
&=&\left(1-\frac{1}{M}\sum_{{\bf S}_0}\left(1-{\rm E}_{{\bf S}_0}\left\{\exp\Big[ - \alpha_p \; \sum_{n=0}^N{\cal
I}\left({\bf S}_n\right)\Big]\right\}\right)\right)^K = \nonumber\\
&=&\exp\Big[-c \sum_{{\bf S}_0} {\rm E}_{{\bf S}_0}\left\{{\cal M}_{\{0\}}\Big(\{{\bf S}_N\}\Big)\right\}\Big], 
\label{eq:7}
\end{eqnarray}
where ${\cal M}_{\{0\}}\Big(\{{\bf S}_N\}\Big)$ is a functional of a given $N$-step random walk trajectory, which is
 defined by
\be
\label{M}
{\cal M}_{\{0\}}\Big(\{{\bf S}_N\}\Big) = 
 \left(1-\exp\Big[ - \alpha_p \; \sum_{n=0}^N{\cal
I}\left({\bf S}_n\right)\Big]\right)
\ee
One notices now that  ${\cal M}_{\{0\}}\Big(\{{\bf S}_N\}\Big)$ counts again  a
number of visits to the origin by a given $N$-step 
random walk trajectory with its starting point being fixed at ${\bf S}_0$. Namely, 
${\cal M}_{\{0\}}\Big(\{{\bf S}_N\}\Big) = 0$ 
for such $N$-step trajectories which never pass through the origin, equals $1-p$ for 
such trajectories which visit the origin
only once, and $1 - p^j$ for those trajectories 
which do it exactly $j$ times. Consequently, the expected value of the functional 
${\cal M}_{\{0\}}\Big(\{{\bf S}_N\}\Big)$, which enters the exponential in Eq.(\ref{eq:7}), 
can be written down as the following polynomial in powers of
the gating probability $p$:
\begin{equation}
{\rm E}_{{\bf S}_0}\left\{{\cal M}_{\{0\}}\Big(\{{\bf S}_N\}\Big)\right\} =
\sum_{j=0}^N\beta_N^{(j)}(0|{\bf S}_0)\Big(1 - p^{j+\delta_{{\bf S}_0,0}}\Big),
\label{eq:nn}
\end{equation}
where $\delta_{{\bf S}_0,0}$ is the Kroneker-delta and  $\beta_N^{(j)}(0|{\bf S}_0)$ is the probability that a 
simple random walk starting from the 
site ${\bf S}_0$ visits 
the origin 
in the first $N$ steps exactly $j$ times \cite{hughes}.  We adhere here to the definition 
 for $\beta_N^{(j)}(0|{\bf S}_0)$ presented in
Ref.\cite{hughes} and  adopt 
the convention
that the initial moment $n = 0$ is regarded as the zeroth visit to the site ${\bf S}_0$.

We turn next to the analysis of the $N$-dependence of the 
integral effective reaction rate for the model I, $Q_N^{(I)}$. 
It follows from Eq.(\ref{eq:nn}) that $Q_N^{(I)}$ is  defined as the polynomial of the form
\be
\label{rate}
Q^{(I)}_N =  \sum_{{\bf S}_0} \sum_{j=0}^N \beta_N^{(j)}(0|{\bf S}_0) \Big(1-p^{j+\delta_{{\bf S}_0,0}}\Big)
\ee
To compute $Q^{(I)}_N$ for any $N$,  we introduce the generating function 
\begin{eqnarray}
Q^{(I)}(\xi)&\equiv&\sum_{N=0}^\infty Q^{(I)}_N\xi^N 
=  (1 - p) \sum_{N =  0}^{\infty}  \beta_N^{(0)}(0|0) \xi^N  + \nonumber\\
&+& 
 \sum_{j=1}^{\infty}\Big(1-p^{j+1}\Big) \sum_{N=1}^\infty \beta_N^{(j)}(0|0)\xi^N +
 \sum_{{\bf S}_0, {\bf S}_0 \neq 0 } \sum_{j=1}^{\infty} \Big(1-p^{j}\Big) \sum_{N=1}^\infty\beta_N^{(j)}(0|{\bf S}_0)\xi^N,
\label{eq:9}
\end{eqnarray}
where we have made use of an evident fact that all $\beta_N^{(j)}(0|{\bf S}_0)$ vanish for $j > N$.

Now, to evaluate  the generating function $Q^{(I)}(\xi)$ explicitly, we calculate 
three different sums entering Eq.(\ref{eq:9}). First of all, we find that
\begin{eqnarray}
\label{u}
(1-p)\sum_{N=0}^\infty\beta_N^{(0)}(0|0)\xi^N=(1-p)\Big\{1+\sum_{N=1}^\infty(1
-\sum_{n=1}^NF_n(0|0))\xi^N\Big\} = \nonumber\\
=(1-p)\Big\{1+\frac{\xi}{1-\xi}-\sum_{n=1}^\infty
F_n(0|0)\sum_{N=n}^\infty\xi^N\Big\}=
\frac{1-p}{1-\xi}(1-F(0|0;\xi)),
\end{eqnarray}
where $F(0|0;\xi)$ is the generating function of $F_n(0|0)$ - the
 average probability that a random walk trajectory starting at the
origin returns to the origin for the first time exactly on the $n$-th
 step \cite{hughes}. Further on, we have 
\begin{eqnarray}
\sum_{j=1}^\infty(1-p^{j+1})\sum_{N=1}^\infty\beta_N^{(j)}(0|0)\xi^N&=&\frac{1}{
1-\xi}\left(1-F(0|0;\xi)\right)\sum_{j=1}^\infty(1-p^{j+1})F(0|0;\xi)^{j}=\nonumber\\
&=&\frac{1-p}{1-\xi}\frac{F(0|0;\xi)(1+p(1-F(0|0;\xi)))}{1-pF(0|0;\xi)},
\end{eqnarray}
and eventually, 
\begin{eqnarray}
\label{z}
&&\sum_{{\bf S}_0\neq0}\sum_{j=1}^\infty(1-p^j)\sum_{N=1}^{\infty}\beta_N^{(j)}(0|{\bf S}_0)\xi^N
=\frac{1-F(0|0;\xi)}{1-\xi}\left[\sum_{j=1}^\infty\Big(1-p^j\Big)
F(0|0;\xi)^{j-1}\right]
\left[\sum_{{\bf S}_0\neq0}F(0|S_0;\xi)\right]=\nonumber\\
&=&\frac{1-F(0|0;\xi)}{1-\xi}\frac{1-p}{\left(1-F(0|0;\xi)\right)\left(1-p
F(0|0;\xi)\right)}(1-F(0|0;\xi))\Big[\frac{1}{1-\xi}-P(0|0;\xi)\Big]=\nonumber\\
&=&\frac{1-p}{1-\xi}\frac{1-F(0|0;\xi)}{1-pF(0|0;\xi)}\Big[\frac{1}{1-\xi}-P(0|0
;\xi)\Big],
\end{eqnarray}
where $P(0|0;\xi)$, as usual, stands for the generating function of the 
 probability $P_N(0|0)$ of having a walker at the origin on the $N$-th
step, provided that the walker started his random walk at the origin.
Consequently, summing up the results in Eqs.(\ref{u}) to (\ref{z}) we obtain 
the following exact expression for the generating function: 
\begin{eqnarray}
Q^{(I)}(\xi)&=& \frac{(1-p)}{(1-\xi)^2}\frac{1-F(0|0;\xi)}{1-pF(0|0;\xi)}=
\frac{1}{(1-\xi)^2}\frac{(1-p)}{p+(1-p)P(0|0;\xi)} = \nonumber\\
&=&  S(\xi) \Big[1 + \frac{p}{(1 - p ) P(0|0;\xi)}\Big]^{-1},
\label{eq:16}
\end{eqnarray}
$S(\xi)$ being the generating
 function of the expected number of distinct sites visited by an $N$-step random walk \cite{hughes},
$S(\xi) = \sum_{N=0}^{\infty} E_0\Big\{{\cal S}\Big(\{{\bf S}_N\}\Big)\Big\} \xi^N$. 
In principle, the integral effective reaction rate for model I valid for any $N$ can be obtained now by  
inverting the
result
in Eq.(\ref{eq:16}), which will require, however, computation of very complex integrals.

We turn now to the analysis of the large-$N$ behavior
 of the reaction rate $Q_N^{(I)}$.  However,  before doing it, it may be expedient
to make first the following observation, which might seem to be quite
surprising at the first glance:

 Namely, the result in  Eq.(\ref{eq:16}) reveals
 that for $\em  recurrent $ random walks 
the leading large-$N$ behavior of the integral 
reaction rate $Q_N^{(I)}$ should be
 $\em independent $ of the gating probability! 
It happens actually because for recurrent walks 
$P(0|0;\xi) \to +\infty$ as $\xi \to 1^-$ ($N \to \infty$), and hence, 
the expression on the rhs of Eq.(\ref{eq:16}) appears to be  independent of $p$. In virtue of the
Tauberian theorem, it implies that  the effective integral
 reaction rate $Q_N^{(I)}$ is independent of $p$ 
when $N \to \infty$. Moreover, the result in Eq.(\ref{eq:16}) shows that
for recurrent walks the leading at $N \to \infty$
behavior of $Q_N^{(I)}$ is defined exactly by the expected number of distinct sites 
visited by an $N$-step random walk.
Consequently, for recurrent walks and $N \to \infty$ 
the target 
survival probability $P_N$ is not influenced by fluctuating gates imposed on the
scavengers and   
has
exactly the same form for reactions 
which are subject to stochastic gating or reactions 
in which the scavengers are always
in the reactive state \cite{blumen,burlatsky,szabo3}.

On the other hand,  such a behavior is 
not counterintuitive and agrees 
with our previous knowledge of the diffusion-limited reactions kinetics.
The point is that imposing a 
fluctuating gate on otherwise perfect scavengers is in a way similar to imposing the constraint that
annihilation of the target by a scavenger
may happen with some finite probability, or at a finite rate prescribed
by certain elementary reaction act constant
$K_{el}$ ($K_{el} < \infty$). Following the 
seminal mean-field analysis of Collins and Kimball \cite{collins} 
(see also Refs.\cite{gleb} and \cite{gleb3} for more details), 
the overall reaction constant 
taking into account both the constraints imposed by the elementary reaction
act (finite $K_{el}$) and the transport limitations
 (in order to react, particles have first to find 
each other in the course of their random motions)  follows
\be
\label{collins}
\Big(\partial Q_N/\partial N\Big)^{-1} = \frac{1}{K_{el}} + \frac{1}{K_{smol}},
\ee
where $K_{smol}$ is the so-called Smoluchowski constant which equals the diffusive current through the surface of an
immobile, perfectly adsorbing sphere.
Now, 
it is well-known (see e.g. Ref.\cite{gleb} for
more discussion) that for low dimensional systems the Smoluchowski constant is not a real constant but rather a
time-dependent coefficient which vanishes as time evolves.
It means that in low dimensions
 random transport of particles 
offers progressively higher resistance with
respect to the overall reaction rate than the
constraints imposed by the 
elementary reaction rate, which results ultimately in the kinetics which is
totally controlled by random transport of particles towards each other and is
independent of $K_{el}$. This is precisely the effect which we observe in case of low-dimensional stochastically-gated
target annihilation problem.  

We note also parenthetically 
that similar effect has been predicted recently for low
dimensional catalytically-activated binary
reactions, in which case the particles' reactivity does not fluctuate in time but is rather a
random function of the space variables \cite{gleb2}. It has been shown here that
long-time kinetics is also insensitive to the concentration of the catalytic sites
which promote reactions between randomly moving $A$ particles and is independent of
$K_{el}$. 
Of course, in higher
dimensional space (such that $d$ is greater than the fractal dimension of the random walk) the
effective reaction rate does depend  on the
 density of catalytic sites and $K_{el}$. Similarly, for stochastically-gated target annihilation
 reactions
$P(0|0;1^-)$ is well defined for $d > 2$, which implies, 
by virtue of Eq.(\ref{eq:16}),  that the leading at $N \to \infty$ terms
in the integral effective reaction
rate should depend on  the gating probability $p$.   

We focus next on the special case of Polya random walks and proceed to 
determine
the long-time behavior of the reaction rate in Eq.(\ref{eq:16}) explicitly, first for one-
and two-dimensional lattices, in which case the Polya walks are recurrent, and then for
$d$-dimensional lattices with $d > 2$, for which the walks are non-recurrent.

\subsection{Polya walks on one-dimensional lattices.}

For Polya random walks on one-dimensional lattices the generating function of the
first-visit probability is 
known exactly and has a particularly simple form (see e.g. Ref.\cite{hughes}), 
$F(0|0;\xi)=1-\sqrt{1-\xi^2}$. Consequently, in this case the generating function $Q^{(I)}(\xi)$
 of the
integral effective reaction rate obeys
\begin{equation}
Q^{(I)}(\xi) =  \frac{(2\xi-1)(1-p)}{(1-\xi)^{3/2}} \frac{\sqrt{1+\xi}}{1-p + p 
\sqrt{1-\xi^2}}
\label{eq:18}
\end{equation}
In the
asymptotical limit $\xi \to 1^-$ (or equivalently, when $N \to \infty$) we find then
from Eq.(\ref{eq:18}) that
\begin{equation}
Q^{(I)}(\xi)= \frac{\sqrt{2}}{(1-\xi)^{3/2}}  -2 \,{\frac {p }{(1-p)\left. (1-\xi\right )}}+
{\mathcal O}\Big(1/\sqrt {\left. (1-\xi\right )}\Big).
\end{equation}
Hence, by virtue of a Tauberian theorem, we have that in the limit 
$N \to \infty$ the effective reaction rate follows
\begin{eqnarray}
\label{ki}
Q^{(I)}_N =  \Big(\frac{8 N}{\pi}\Big)^{1/2}    - 2 \, \frac{p}{1-p}
+  {\mathcal O}\Big({
\frac {1}{\sqrt {N}}}\Big),
\end{eqnarray}
i.e., as we have already remarked, the leading behavior as $N \to \infty$ in
 case of the target annihilation problem with
stochastic gates imposed on the scavengers is
 exactly the same as in case of its ungated counterpart, Eq.(\ref{visits}). The first
correction term, however, does depend on the
 gating probability $p$ and diverges when $p \to 1$, i.e.
 in the limit when scavengers
are being completely inert
 with respect to the reaction. Simple comparison of the first two terms in Eq.(\ref{ki}) 
shows that the universal, $p$-independent behavior is established
 when $N$ exceeds certain cross-over value $N^{*}$,
such that $N^{*} \approx \pi p^2/2 (1 - p)^2$. Note also 
that similar behavior has been predicted earlier 
in
Ref.\cite{ber1} within the framework of a continuous-space description.

\subsection{Polya walks on two-dimensional lattices.}

The generating function $P(0,0|\xi)$ is not known explicitly for Polya walks on
two-dimensional lattices. However, its asymptotical behavior as $\xi \to 1^-$ (or,
equivalently, when $N \to \infty$) is well documented (see e.g. \cite{hughes}) and is given by
\be
P(0|0;\xi)=\frac{1}{\pi C_2} \ln\Big(\frac{K}{1-\xi}\Big)(1+ {\mathcal O}( 1-\xi)),
\ee
where
the constant $C_2$ has been defined in the text after Eq.(\ref{visits}), while
the constant $K$ equals $4$, $8$ and $12$ for hexagonal, square and triangular lattices, respectively. 

From the latter equation we find than that the leading asymptotical
behavior of $Q^{(I)}(\xi)$ as $\xi \to 1^-$ follows
\begin{eqnarray}
Q^{(I)}(\xi) &=& - \frac{\pi C_2}{ (1-\xi)^{2} \ln(1-\xi) }
- \nonumber\\
&-& \pi^{2}\Big(p+\frac{(1-p)\ln (K)}{\pi C_2}
\Big) \frac{C_{2}^2}{(1-p)(1-\xi)^{2} \ln^2(1-\xi)}  + 
{\mathcal O}\Big(\frac{1}{\ln^3(1-\xi) (1-\xi)^{2}}\Big)
\end{eqnarray}
Hence, by
applying the Tauberian theorem 
we find that for two-dimensional target annihilation with stochastically-gated scavengers
the integral effective reaction rate obeys
\begin{equation}
Q^{(I)}_N = \pi C_2 \, \frac {  N}{\ln (N)}+ \pi \, C_2  
\Big(1 - \gamma - \ln(K) - 
p \pi/C_2 (1-p)\Big) \, \frac {N}{\ln^2(N)} +
 {\mathcal O}\Big(\frac{N}{\ln^3(N)}\Big),
\label{long}
\end{equation}
where $\gamma$ denotes the Euler constant. 
Note that again, in accord with our earlier prediction, the leading large-$N$ behavior appears to be
independent of the gating probability and proceeds exactly in the same way as for the ungated target problem.
This long-time regime can be observed, however, at considerably longer times than that for the one-dimensional
systems; on comparing the first two terms on the rhs of Eq.(\ref{long}) we infer that the corresponding
cross-over time $N^{*}$ is given by
\be
N^{*} \approx \exp\Big[\frac{\pi C_2 p}{1 - p}\Big],
\ee
i.e. is $\em exponentially$ large when $p \to 1$, while in  one-dimensional systems
 this dependence is only
algebraic.

\subsection{Polya walks on $d$-dimesional lattices, $d > 2$}

Lastly, we turn to the case of recurrent Polya walks, which case is realized, namely, for lattices with spatial
dimension $d > 2$. Here the probability $R$ of eventual return to the origin is finite, and
consequently, we find from Eq.(\ref{eq:16}) that
\begin{equation}
Q^{(I)}(\xi) = \frac{1-p}{(1-p)P(0|0;1^-)+p}\;\frac{1}{(1-\xi)^2}  + {\mathcal O}\Big(\frac{1}{(1-\xi)^{3/2}}\Big),
\label{eq:25}
\end{equation}
which yields, in the large-$N$ limit, the following result
\begin{equation}
Q^{(I)}_N = \frac{1-p}{(1-p)P(0|0;1^-)+p}\;N +  {\mathcal O}\Big(\sqrt{N}\Big)
\label{eq:26}
\end{equation}
Hence, for lattices with $d > 2$ the decay of the survival probability is purely exponential in all dimensions.
Note also that the exact result in Eq.(\ref{eq:26}) confirms in a way the
mean-field result by Collins and Kimball,  Eq.(\ref{collins}); as a matter of fact, it appears that 
Eq.(\ref{eq:26}) can be cast
exactly into the form of Eq.(\ref{collins})
 if we set $K_{el} = (1-p)/p$ and 
$K_{smol} = 1/P(0|0;1^-)$.  Note also that our Eq.(\ref{eq:26}) confirms the conclusion of Szabo et al
\cite{szabo1} concerning the possibility of the calculation of the steady-state stochastically-gated rate
constant in terms of an appropriately defined ungated model.

\section{Model II: An immobile fluctuating target and randomly moving ungated scavengers.}

We turn next to the survival of a $\em stochastically$-gated, immobile  $A$  particle, - a target,
 in the presence of ungated scavengers $B$, which perform independent
random walks on a $d$-dimensional lattice. 
For this model the indicator function of the reaction
event can be written down as follows 
\begin{eqnarray}
\Psi_N=\lim_{\beta\to\infty}\exp\left[-\beta\sum_{n=0}^N\eta_n\;\sum_{k=1}^K{\cal
I}\left({\bf S}_n^{(k)}\right)\right],
\label{eq:41}
\end{eqnarray}
where $\eta_n$ is the indicator variable of the gate imposed on the target, while 
${\bf S}_n^{(k)}$ defines the lattice positions of the $k$-th scavenger at 
the $n$-th step, $n=0,1, \ldots ,N$.
We again suppose  
 that 
the target reactivity assumes at random
two values - $1$ and $0$ with probabilities $1 - p$ and $p$, respectively. In the state 
$\eta_n = 1$  the target is accessible for reaction and can be annihilated by any of the
scavengers arriving at the origin, while in the state $\eta_n = 0$ reaction can not take place.

Averaging first $\Psi_N$ in Eq.(\ref{eq:41}) with respect to the fluctuations of the reactivity, we get
\begin{eqnarray}
\overline{\Psi_N}&=&\overline{\prod_{n=0}^N\lim_{\beta\to\infty}\exp\left[-\beta\eta_n \;\sum_{k=1}^K{\cal
I}\left({\bf S}_n^{(k)}\right)\right]}\nonumber\\
&=&\prod_{n=0}^N\left\{(1-p)\lim_{\beta\to\infty}\exp\left[-\beta\sum_{k=1}^K{\cal
I}\left({\bf S}_n^{(k)}\right)\right]+p\right\}
\label{eq:42}
\end{eqnarray}
Furter on, noticing that 
\begin{equation}
\label{eq:43}
(1-p)\lim_{\beta\to\infty}\exp\left[-\beta\sum_{k=1}^K{\cal 
I}\left({\bf S}_n^{(k)}\right)\right]+p= \left\{\begin{array}{ll}
1,   \;   \mbox{$\sum_{k=1}^K{\cal
I}\left({\bf S}_n^{(k)}\right)=0$},\\
p,  \;    \mbox{$\sum_{k=1}^K{\cal
I}\left({\bf S}_n^{(k)}\right) > 0$,}
\end{array}
\right.
\end{equation}
and hence, rewriting this expression as
\begin{equation}
(1-p)\lim_{\beta\to\infty}\exp\left[-\beta\sum_{k=1}^K{\cal
I}\left({\bf S}_n^{(k)}\right)\right]+p=\exp\Big[ -\alpha_p \Big( 1 -  {\cal
I}\left( \sum_{k=1}^K{\cal
I}\left({\bf S}_n^{(k)}\right)\right)\Big)\Big],
\label{eq:44}
\end{equation}
we find that the indicator function 
of the reaction event, averaged over the fluctuations of the target reactivity, attains the
form
\be
\label{blin}
\overline{\Psi_N}=  \exp\Big[ -\alpha_p \; {\cal N}^*\left(\{{\bf S}_N^{(k)}\}\right)\Big],
\ee
where ${\cal N^*}\left(\{{\bf S}_N^{(k)}\}\right)$ is given by
\be
{\cal N^*}\left(\{{\bf S}_N^{(k)}\}\right) =  \sum_{n=0}^N \Big( 1 -  {\cal
I}\left( \sum_{k=1}^K{\cal
I}\left({\bf S}_n^{(k)}\right)\right)\Big)
\label{eq:45}
\ee
Note now that the functional $\Big(1-{\cal I}\left(\sum_{k=1}^K{\cal
I}\left({\bf S}_n^{(k)}\right)\right)\Big)$ measures
 the occupancy of the origin at time moment $n$.
 It equals zero if none of $K$ walkers is present at the origin at the time moment $n$ and
equals one if one or several scavengers appear at the origin at the $n$-th step. In this regard, 
${\cal N^*}\Big(\{{\bf S}_N^{(k)}\}\Big)$ is similar to the earlier defined functional 
${\cal N}\left(\{{\bf S}_N^{(k)}\}\right)$ appearing in the analysis of the model I. 
Important difference is, 
however, that a simultaneous  visit of the origin by several walkers  is  counted
 as a single visit, and consequently, 
${\cal N^*}\left(\{{\bf S}_N^{(k)}\}\right)$ describes $\em collective$ behavior of all $K$ walkers,
 which can not be 
factorized, as it appears in the model I. 
This substantial distinction between the models involving
ungated and gated targets has been noticed already by Szabo 
et al in Ref.\cite{szabo2} (see also Ref.\cite{ber1} for more
details), who stated that
 the crucial difference between the case when
the gates are imposed on $B$s or on the $A$ particle is that
in the latter case the "switching of the $A$ from the reactive conformation to a
non-reactive one is felt simulateneously by all scavengers". 
It means, in particular,  that if we define the 
functional ${\cal N^*}\left(\{{\bf S}_N^{(k)}\}\right)$ 
using the "directed" polymers
representation in Fig.4, then it would count all sites on the $n$-axis visited simultaneously by two,
 three and etc walkers
as singly visited sites. In this regard, ${\cal N^*}\left(\{{\bf S}_N^{(k)}\}\right)$ determines the number of $\em distinct$
visits to the origin by $K$ independent walkers. 

We notice next that an 
 upper bound on the integral effective reaction rate $Q^{(II)}_N$ for the model II
can be found very straightforwardly. To do this, it suffices merely to observe that
\be
1-{\cal I}\left(\sum_{k=1}^K{\cal
I}\left({\bf S}_n^{(k)}\right)\right) \leq \sum_{k=1}^K{\cal
I}\left({\bf S}_n^{(k)}\right),
\label{m}
\ee
and hence, that ${\cal N^*}\left(\{{\bf S}_N^{(k)}\right) \leq 
{\cal N}\left(\{{\bf S}_N^{(k)}\}\right)$. This implies, in turn, that the
 survival probability $P_N$ for the model II is greater than 
the survival probability obtained for the model I, and the integral effective reaction rate $Q^{(II)}_N$ obeys
\be
\label{upper}
Q^{(II)}_N \leq Q^{(I)}_N,
\ee
which inequality sets a rigorous upper bound on $Q^{(II)}_N$. Note that the inequality in Eq.(\ref{upper}) has
been
established earlier using different type of arguments in Ref.\cite{ber2}. 

It may be worthy to remark that the inequality in Eq.(\ref{m}) holds as an equality when all ${\bf S}_n^{(k)}$ are
different at a given $n$, which happens, namely, when the scavengers do obey a hard-core exclusion and none 
two scavengers can 
simultaneously occupy the same lattice site. Given that the scavegers are indistinguishable, we may thus expect that
for the model II with hard-core scavengers the $A$ particle survival probability will be determined exactly by
Eqs.(\ref{ki}) to (\ref{eq:26}) 
at sufficiently
large times.

We proceed next to calculation of the upper bound on the survival probability for the model II, which requires
 a bit more complicated analysis. To do this, 
we return to the indicator function of the reaction event in Eq.(\ref{eq:42}) and perform first 
averaging  with respect to  the trajectories and the initial positions of the
scavengers. The steps involved in the averaging procedure in this case are as follows:
\begin{eqnarray}
<\Psi_N>&=&<\prod_{k=1}^K\lim_{\beta\to\infty}\exp\Big\{-\beta\sum_{n=0}^N\eta_n \; {\cal
I}\Big({\bf S}_n^{(k)}\Big)\Big\}>
=\prod_{k=1}^K<\lim_{\beta\to\infty}\exp\Big\{-\beta\sum_{n=0}^N\eta_n \; {\cal
I}\Big({\bf S}_n^{(k)}\Big)\Big\}>=\nonumber\\
&=&\prod_{k=1}^K\Big(\frac{1}{M}\sum_{{\bf S}_0}{\rm
E}_{{\bf S}_0}\Big\{\lim_{\beta\to\infty}\exp\Big[-\beta \; \sum_{n=0}^N\eta_n \; {\cal I}\Big({\bf
S}_n^{(k)}\Big)\Big]\Big\}\Big)=\nonumber\\
&=&\Big(\frac{1}{M}\sum_{{\bf S}_0}{\rm
Prob}\Big(\eta_n \; {\cal I}\Big({\bf
S}_n\Big)=0  \; 
\text{for any } n \in [0,N]|{\bf S}_0\Big)\Big)^K,
\label{eq:480}
\end{eqnarray}
where ${\rm
Prob}\Big(\eta_n \; {\cal I}\Big({\bf
S}_n\Big)=0  \; 
\text{for any } n \in [0,N]|{\bf
S}_0\Big)$  is the probability that an $N$-step random walk, 
starting at ${\bf S}_0$ and
 characterized by internal two-state variable $\eta_n$,  
has never visited the origin being in 
the reactive state $\eta_n = 1$. Next, turning to the infinite-space limit, we find
\begin{eqnarray}
<\Psi_N>&=&\Big(1 - \frac{1}{M}\sum_{{\bf S}_0}\Big(1 - {\rm
Prob}\Big(\eta_n \; {\cal I}\Big({\bf
S}_n\Big)=0  \; 
\text{for any } \; n \in [0,N]|{\bf
S}_0\Big)\Big)\Big)^K=\nonumber\\
&=&\exp\Big[ - c \; \sum_{{\bf S}_0}\Big(1 - {\rm Prob}\Big(\eta_n \; {\cal I}\Big({\bf
S}_n\Big)=0  \; 
\text{for any } n \in [0,N]|{\bf S}_0 \Big) \Big) \Big]
\label{eq:48}
\end{eqnarray}
Evidently, 
\be
\Big(1 - {\rm Prob}\Big(\eta_n \; {\cal I}\Big({\bf
S}_n\Big)=0  \; 
\text{for any } n \in [0,N]|{\bf S}_0 \Big) \Big) =  {\rm Prob}\Big(\sum_{n=0}^N \eta_n \; {\cal I}\Big({\bf
S}_n\Big) \geq 1|{\bf S}_0 \Big),
\ee
where  ${\rm Prob}\Big(\sum_{n=0}^N \eta_n \; {\cal I}\Big({\bf
S}_n\Big) \geq 1|{\bf S}_0 \Big)$ is the probability  that an $N$-step random walk starting 
at the site ${\bf S}_0$ has at least once visited the origin being in the
reactive state. 

Note that ${\rm Prob}\Big(\sum_{n=0}^N \eta_n \; {\cal I}\Big({\bf
S}_n\Big) \geq 1|{\bf S}_0 \Big)$ is not constrained in the sense that it provides no information
at which of the visits to the origin precisely the
reactive state has appeared; that is, the particle could visit the origin many times until it
arrived eventually being in the reactive state. Having this in mind,   
 we now realize  
how the 
sum in the exponent in the last line of Eq.(\ref{eq:48})
 can be bounded from below, which will result in
the desired upper bound on the target survival probability in the
model II. To do this, we will proceed as follows: Suppose first that the starting point of the trajectory
$\{ {\bf S}_N\}$ is not the origin. 
Then, we notice that for ${\bf S}_0 \neq 0$, one has
\begin{equation}
\label{dina}
{\rm Prob}\Big(\sum_{n=0}^N \eta_n \; {\cal I}\Big({\bf
S}_n\Big) \geq 1|{\bf S}_0 \Big)  \geq \sum_{n=0}^N  {\rm
Prob}\left(\eta_n \; {\cal I}\Big({\bf
S}_n\Big) = 1  \; \text{and} \sum_{l=0}^{n-1} \eta_l \; {\cal I}\Big({\bf
S}_l\Big) =0|{\bf
S}_0 \right),
\end{equation}
where ${\rm
Prob}\left(\eta_n \; {\cal I}\Big({\bf
S}_n\Big) = 1  \; \text{and} \sum_{l=0}^{n-1} \eta_l \; {\cal I}\Big({\bf
S}_l\Big) =0|{\bf
S}_0 \right)$  stands for the probability that the origin has been
 visited for the first time exactly at the $n$-th step (has not been visited prior to the $n$-th step)
and the particle at this very step was in the reactive state, 
i.e. such that $\eta_n = 1$. Hence, the rhs of Eq.(\ref{dina}) is the  
constrained probability that within
an $N$-step walk starting at the site ${\bf S}_0 \neq 0$ the particle happened to be in the reactive
state at its first visit to the origin. 
Summing next both sides of the inequality in Eq.(\ref{dina}) over all starting points, we obtain 
\begin{eqnarray}
& &\sum_{{\bf S}_0} {\rm Prob}\Big(\sum_{n=0}^N \eta_n \; {\cal I}\Big({\bf
S}_n\Big) \geq 1|{\bf S}_0 \Big)  \geq \sum_{{\bf S}_0, {\bf S}_0 \neq 0} 
{\rm Prob}\Big(\sum_{n=0}^N \eta_n \; {\cal I}\Big({\bf
S}_n\Big) \geq 1|{\bf S}_0 \Big)   \geq \nonumber\\
&&\geq \sum_{{\bf S}_0, {\bf S}_0 \neq 0} \sum_{n=0}^N  {\rm
Prob}\left(\eta_n \; {\cal I}\Big({\bf
S}_n\Big) = 1 \; \text{and} \sum_{l=0}^{n-1} \eta_l \; {\cal I}\Big({\bf
S}_l\Big) =0 |{\bf
S}_0 \right) = \sum_{{\bf S}_0, {\bf S}_0 \neq 0}\sum_{n=0}^N \eta_n \;
 F_n(0|{\bf S}_0) =\nonumber\\
&=& \sum_{n=0}^N\eta_n\sum_{{\bf S}_0, {\bf S}_0 \neq 0}F_n(0|{\bf S}_0)
=\sum_{n=0}^N\eta_n\sum_{{\bf S}_0}F_n({\bf S}_0|0)=
\sum_{n=0}^N\eta_n E_0\Big\{\Delta\Big(\{{\bf S}_n\}\Big)\Big\},
\label{eq:51}
\end{eqnarray}
where $\Delta\Big(\{{\bf S}_n\}\Big)$ is an auxiliary random variable,
 defined to be the number of $\em virgin$ sites
visited on the $n$-th step by a 
given particle trajectory $\{{\bf S}_n\}$ \cite{hughes}, 
\be
\label{virgin}
\Delta\Big(\{{\bf S}_n\}\Big)   = {\cal S}\Big(\{{\bf S}_n\}\Big) 
- {\cal S}\Big(\{{\bf S}_{n-1}\}\Big)  
\ee
Consequently, 
we  can bound the rhs of Eq.(\ref{eq:48}) as
\begin{eqnarray}
P_N&\leq&\overline{\exp\left[- \; c \; 
 \sum_{n=0}^N\eta_n \; E_0\Big\{\Delta\Big(\{{\bf S}_n\}\Big)\Big\} \right]}
 = \prod_{n=0}^N \overline{\exp\left[- \; c \;  \eta_n \; 
 E_0\Big\{\Delta\Big(\{{\bf S}_n\}\Big)\Big\}
\right]}=\nonumber\\
&=&\prod_{n=0}^N\left\{
(1-p) \exp[- \; c \; E_0\Big\{\Delta\Big(\{{\bf S}_n\}\Big)\Big\} ] + p \right\}
\label{eq:500}
\end{eqnarray}
Further on, following the Dvoretzky-Erd\"os Lemma 
$E_0\Big\{\Delta\Big(\{{\bf S}_n\}\Big)\Big\}$ is a monotonic 
decreasing sequence 
of time $n$ (see, e.g. \cite{hughes}), i.e. 
\be
1 \geq E_0\Big\{\Delta\Big(\{{\bf S}_1\}\Big)\Big\} 
\geq E_0\Big\{\Delta\Big(\{{\bf S}_2\}\Big)\Big\} \geq \; \ldots \geq 
 E_0\Big\{\Delta\Big(\{{\bf S}_n\}\Big)\Big\} \geq \; \ldots \geq E_0\Big\{\Delta\Big(\{{\bf
S}_N\}\Big)\Big\},
\ee
such that
\be
lim_{n \to \infty} E_0\Big\{\Delta\Big(\{{\bf S}_n\}\Big)\Big\}  = \frac{1}{P(0|0;1^-)},
\ee
we can majorize the terms in the curly brackets on the rhs of Eq.(\ref{eq:500}) by replacing
$E_0\Big\{\Delta\Big(\{{\bf S}_n\}\Big)\Big\}$ by its minimal value 
$E_0\Big\{\Delta\Big(\{{\bf S}_N\}\Big)\Big\}$. 
Enhancing in such a way the bound in Eq.(\ref{eq:500}), we have that 
\begin{eqnarray}
P_N&\leq& 
\prod_{n=0}^N \left\{
(1-p) \exp[- \; c \; E_0\Big\{\Delta\Big(\{{\bf S}_N\}\Big)\Big\} ] + p \right\} = \nonumber\\
&=& \left\{
(1-p) \exp[- \; c \; E_0\Big\{\Delta\Big(\{{\bf S}_N\}\Big)\Big\} ] + p \right\}^{N+1} = \nonumber\\
&=& \exp\Big[- \; (N + 1) \; \ln\Big(1/\Big( 
(1-p) \exp[- \; c \; E_0\Big\{\Delta\Big(\{{\bf S}_N\}\Big)\Big\} ] + p  \Big) \Big)\Big]
\label{eq:52}
\end{eqnarray}
and hence, the integral effective rate constant obeys
\begin{equation}
Q^{(II)}_N \geq  \frac{(N + 1)}{c} \; \ln\Big(1/\Big( 
(1-p) \exp[- \; c \; E_0\Big\{\Delta\Big(\{{\bf S}_N\}\Big)\Big\} ] + p  \Big) \Big)
\label{eq:53}
\end{equation}
For $d$-dimesnional Polya walks, in particular, we find from Eq.(\ref{eq:53}) the following
 explicit lower bounds
on the integral effective rate constant in the model II:
\begin{eqnarray}
&d&=1 \; \; \;  Q^{(II)}_N \geq   \; (1 - p) \; \Big(\frac{2 N}{\pi }\Big)^{1/2} \Big(1 + {\mathcal
O}\Big(1/\sqrt{N}\Big) \Big), \nonumber \\
&d&=2 \; \; \;  Q^{(II)}_N \geq   \; (1 - p) \;  \frac{ \pi C_2 N}{\ln(N)} \Big(1 +  {\mathcal
O}\Big(1/\ln(N)\Big) \Big), \nonumber \\
&d&> 2 \; \; \; Q^{(II)}_N \geq  \frac{N}{c} \; \ln\Big(1/\Big( 
(1-p) \exp[- \; c / P(0|0;1^-) ] + p \Big) \Big) \Big(1 +  {\mathcal
O}\Big(1/\sqrt{N}\Big)\Big),
\label{eq:54}
\end{eqnarray}
which hold in the limit $N \to \infty$.

On comparing the results in
Eqs.(\ref{upper}) and (\ref{eq:54}), 
we notice  that both lower and upper bounds display the same $N$-dependence, 
but differ slightly in numerical factors. This means, in
turn, that the $N$-dependence of the integral effective 
reaction rate $Q^{(II)}_N$ is determined here exactly
 by Eqs.(\ref{upper}) and (\ref{eq:54}). Consequently, at sufficiently large times
 the decay laws in the
models I and II of gated target annihilation are essentially the same 
(up to a possible difference in
characteristic decay times), and  
coincide with the decay law predicted for the ungated model of Section II.

Note also that the time evolution of the $A$ particle survival probability 
for the model II defined in a one-dimensional continuum 
has been considered earlier 
in Ref.\cite{ber1}. Within the framework of the 
heuristic Smoluchowski-type approach, it has been predicted  
that the long-time decay of $P_N$ should follow the decay law 
in the first line in Eq.(\ref{visits}),
i.e. should proceed at long times exactly in the same fashion as that for the model I and consequently,   
should be essentially the same as in the ungated target problem. 
While intuitively such a behavior
seems to be quite plausible for low-dimensional systems (see the discussion following the Collins-Kimball
result in Eq.(\ref{collins})) and, as a matter of fact,
 agrees with our prediction in
Eq.(\ref{upper}), one still can not, in principle,  
rule out the possibility that the integral effective reaction rates for models I and II may
have different numerical factors even in low dimensions.
 The point is that the Smoluchowski approach in
Ref.\cite{ber1}, which 
 is a continuous-space counterpart of the Rosenstock approximation, allows to determine
 here only a lower bound on
the target survival probability and thus can not produce exact numerical factors.

\section{Model III: 
An ungated, mobile $A$ particle and immobile, randomly placed, stochastically-gated traps $B$.}

We turn next to the case of  
stochastically-gated trapping reactions, focusing first on the situation involving
 an ungated 
$A$ particle, which performs
 a discrete-time, homogeneous random walk on  a $d$-dimensional lattice
starting from the origin at $n = 0$, in
the presence
of immobile, randomly placed, stochastically-gated $B$ particles. 
The properties of the gates are the same as defined in the model I.

For this model, the indicator function of the reaction event can be written down
as follows
\begin{eqnarray}
\Psi_N&=&\lim_{\beta\to\infty}\exp\left[-\beta\sum_{n=0}^N\sum_{k=1}^K{\cal I}\left({\bf
r}_n-{\bf S}^{(k)}\right)\eta_n^{(k)}\right]\nonumber\\
&=&\prod_{n=0}^N\prod_{k=1}^K\lim_{\beta\to\infty}\exp\left[-\beta{\cal I}\left({\bf
r}_n-{\bf S}^{(k)}\right)\eta_n^{(k)}\right],
\label{eq:27}
\end{eqnarray}
where ${\bf S}^{(k)}$ are $d$-dimensional lattice vectors denoting positions of $K$ traps $B$, while ${\bf
r}_n$ defines the lattice position of the $A$ particle at the $n$-th step. 

Averaging first over the reactivity of the traps, we readily find
\begin{eqnarray}
\label{eq:28}
\overline{\Psi_N}&=&\prod_{n=0}^N\prod_{k=1}^K\overline{\lim_{\beta\to\infty}\exp\left[-\beta \; {\cal I}\left({\bf
r}_n-{\bf S}^{(k)}\right) \; \eta_n^{(k)}\right]}=\nonumber\\
&=&\prod_{n=0}^N\prod_{k=1}^K \Big\{ (1-p)\lim_{\beta\to\infty} \exp\Big[-\beta \; {\cal I}\left({\bf
r}_n-{\bf S}^{(k)}\right) \Big]+p \Big\}
\end{eqnarray}
Further on, noticing that 
\begin{equation}
\label{eq:29}
(1-p) \lim_{\beta\to\infty} \exp\left[-\beta \; {\cal I}\left({\bf
r}_n-{\bf S}^{(k)}\right)\right]+p = \left\{\begin{array}{ll}
p,   \;   \mbox{ ${\bf r}_n = {\bf S}^{(k)}$},\\
1,  \;    \mbox{${\bf r}_n \neq {\bf S}^{(k)}$},
\end{array}
\right.
\end{equation}
and hence, rewriting this expression as
\begin{equation}
\label{eq:30}
(1-p)\lim_{\beta\to\infty}\exp\Big[-\beta \; {\cal I}\Big({\bf
r}_n-{\bf S}^k\Big)\Big]+p=\exp\Big[ - \alpha_p {\cal I}\Big({\bf
r}_n-{\bf S}^k\Big)\Big],
\end{equation}
we have that the indicator function of
 the reaction event, averaged over the reactivity fluctuations, takes the
following form:
\begin{equation}
\overline{\Psi_N}=\exp\Big[ - \alpha_p \; \sum_{n=0}^N\sum_{k=1}^K{\cal I}\left({\bf
r}_n-{\bf S}^{(k)}\right)\Big]
\label{eq:31}
\end{equation}
Note, that the averaged indicator function $\overline{\Psi_N}$ is now
an exponential of the factor $\alpha_p$ times the number of
times a given random walk trajectory 
starting at $n = 0$ at the origin visits a 
given  array of lattice sites 
$\{{\bf S}^{(k)}\}$, i.e. can be thought off as the moment generating function
of  the "residence time" of 
a single random walker on a  given  array $\{{\bf S}^{(k)}\}$.
From a different perspective, 
$\Psi_N$ can be viewed as the partition function of a phantom polymer chain on a
lattice with randomly placed energetic barriers of finite height;
 the limit $\alpha_p  \to \infty$, ($p \to 0$),
 corresponding then to the case of
randomly placed, impenetrable obstacles \cite{nechaev}.

Now, the double average over the trajectories of the $A$ particle and over
 the  positions of the
traps  can be written down as follows
\begin{eqnarray}
P_N=<\overline{\Psi_N}>&=&E_0\Big\{<\prod_{k=1}^K \exp\Big[ - \alpha_p \; \sum_{n=0}^N {\cal
I}\left({\bf
r}_n-{\bf S}^{(k)}\right)\Big]>\Big\} = \nonumber\\
&=&E_0 \Big\{ \prod_{k=1}^K <\exp\Big[ - \alpha_p \; \sum_{n=0}^N {\cal I}
\Big({\bf r}_n-{\bf S}^{(k)}\Big)\Big]> \Big\} =\nonumber\\
&=&E_0 \Big\{ \Big( \frac{1}{M} \sum_{{\bf S}}\exp\Big[ - \alpha_p \; \sum_{n=0}^N{\cal
I}\Big({\bf
r}_n-{\bf S}\Big)\Big]\Big)^K\Big\}
\end{eqnarray}
Turning next to the infinite-space limit, i.e. letting $M,K \to \infty$, while keeping their ratio fixed,
we have that
\begin{eqnarray}
P_N
&=&E_0 \Big\{ \Big(1-\frac{1}{M}\sum_{{\bf S}}\Big(1-\exp\Big[- \alpha_p \;
\sum_{n=0}^N{\cal 
I}\Big({\bf
r}_n-{\bf S}\Big)\Big]\Big)\Big)^K\Big\}=\nonumber\\
&=&E_0\Big\{\exp\left[-c\sum_{{\bf S}} {\cal M}_{\{{\bf S}\}}\Big(\{{\bf
r}_N\}\Big)\right]\Big\},
\label{eq:32}
\end{eqnarray} 
where ${\cal M}_{\{{\bf S} \}}\Big(\{{\bf
r}_N\}\Big)$ is defined by
\be
\label{MR}
{\cal M}_{\{{\bf S} \}}\Big({\bf
r}_n\Big) = \left(1-\exp\Big[- \alpha_p \; \sum_{n=0}^N{\cal
I}\left({\bf
r}_n-{\bf S}\right)\Big]\right)
\ee
Note that ${\cal M}_{\{{\bf S} \}}\Big(\{{\bf
r}_N\}\Big)$
is quite similar to the functional defined in Eq.(\ref{M}) 
with the only minor difference being that the latter is associated 
with the multiple visits to some given
 site  ${\bf S}$; namely,   ${\cal M}_{\{{\bf S} \}}\Big(\{{\bf
r}_N\}\Big) = 1 - p^j$ if the site ${\bf S}$ is visited exactly $j$ 
times by an $N$-step walk starting at the site ${\bf r}_0 = 0$.
Consequently, the sum $\sum_{{\bf S}} {\cal M}_{\{{\bf S} \}}\Big(\{{\bf
r}_N\}\Big)$ probes the occupancy of the sites visited by a given realization of a 
random walk trajectory (see Fig.5).

However, an important difference with the previously considered model,
 which makes the computation of
$P_N$ for the model III to be a fairly complex problem, is that here 
we have to deal with a moment generating
function of ${\cal M}_{\{{\bf S} \}}\Big(\{{\bf
r}_N\}\Big)$, compared to a much easier problem of computation of an expected
value of ${\cal M}_{\{{\bf S} \}}\Big(\{{\bf
r}_N\}\Big)$, 
encountered in the model I. Below we consider first predictions based on 
some approximate approach - the 
Rosenstock approximation, and next derive rigorous lower and upper bounds,
 which both have the same
dependence on the time $N$ but differ in the prefactors. 

\subsection{Rosenstock approximation. Decay pattern at intermediate times $N$.}

We start our analysis of the $N$-dependence of the survival probability in
 Eq.(\ref{eq:32}) by considering
first the predictions of the
Rosenstock approximation. 
Applying the Jensen inequality, we may 
bound $P_N$ in Eq.(\ref{eq:32}) as follows
\be
\label{rosen}
P_N \geq \exp\Big[ - c E_0\Big\{   \sum_{{\bf S}} {\cal M}_{\{{\bf S}\}}\Big(\{{\bf
r}_N\}\Big)\Big\} \Big]
\ee
To proceed further, we note that 
the sum 
$\sum_{{\bf S}} {\cal M}_{\{{\bf S} \}}\Big(\{{\bf
r}_N\}\Big)$ can be written down formally (see Fig.5) as the 
following polynomial with random coefficients 
\be
\label{zu}
\sum_{{\bf S}} {\cal M}_{\{{\bf S}\}}\Big(\{{\bf
r}_N\}\Big) = \sum_{j = 1}^{N} {\cal V}^{(j)}\Big(\{{\bf
r}_N\}\Big) \; \Big(1 - p^j\Big) = {\cal S}\Big(\{{\bf
r}_N\}\Big) -  \sum_{j = 1}^{N} {\cal V}^{(j)}\Big(\{{\bf
r}_N\}\Big) \; p^j,
\ee
where each ${\cal V}^{(j)}\Big(\{{\bf
r}_N\}\Big)$, $j = 1, ... , N$, 
 is a random variable, which equals the number of sites in a given $N$-step random walk trajectory $\{{\bf
r}_N\}$  being visited exactly $j$ times. Note that ${\cal V}^{(j)}\Big(\{{\bf
r}_N\}\Big)$ are not independent; this can be readily seen 
if one notices that the combination $\sum_{j = 1}^N j {\cal V}^{(j)}\Big(\{{\bf
r}_N\}\Big)$ is a non-fluctuating quantity and equals the total number of
sites visited by an $N$-step walk, i.e. $N$. Lastly, 
the random function ${\cal S}\Big(\{{\bf
r}_N\}\Big)$ in Eq.(\ref{zu}) denotes, as previously defined, the number of $\em distinct$
sites visited by an
 $N$-step random walk trajectory $\{{\bf
r}_N\}$, ${\cal S}\Big(\{{\bf
r}_N\}\Big) = \sum_{j = 1}^{N} {\cal V}^{(j)}\Big(\{{\bf
r}_N\}\Big)$. 

Consequently, we find that Eq.(\ref{rosen}) attains the following form
\be
\label{r}
P_N \geq \exp\Big[ - c E_0\Big\{ {\cal S}\Big(\{{\bf
r}_N\}\Big)\Big\} + c  \sum_{j = 1}^{N}  E_0\Big\{ {\cal V}^{(j)}\Big(\{{\bf r}_N\}\Big) \Big\} 
\; p^j  \Big]
\ee
The averaging  in the exponent in Eq.(\ref{r}) can be performed directly
using the results obtained in \cite{hughes} for the
generating function of the expectation $E_0\Big\{{\cal V}^{(j)}\Big(\{{\bf
r}_N\}\Big)\Big\}$.  On the other hand, it is evident that for homogeneous random walks
\be
\label{ap}
E_0\Big\{\sum_{{\bf S}} {\cal M}_{\{{\bf S}\}}\Big(\{{\bf
r}_N\}\Big)\Big\} \equiv \sum_{{\bf r}_0} E_{{\bf r}_0}\Big\{{\cal M}_{\{0\}}\Big(\{{\bf
r}_N\}\Big)\Big\},
\ee
which implies, in turn, that in terms of the Rosenstock approximation 
the integral effective reaction rate
 for the model III, $Q^{(III)}_N$,  
 has exactly the same form as that obtained for the model I. As a matter of fact, this does not
seem to be surprising since the mean-field-type Rosenstock approximation is not sensitive to the
fact which of the reactive species precisely 
is mobile and which is fixed (see Eqs.(\ref{rosen}) and
(\ref{ap})). However, a profound difference between these two models does exist and below we will show
that the large-$N$ decay of $P_N$ proceeds slower than it is predicted by Eqs.(39),(42) and (45).

\subsection{Large-$N$ decay of the survival probability.}  

First of all, we note that in virtue of Eq.(\ref{zu}) and of an evident observation that
 for any given random walk trajectory $\{{\bf
r}_N\}$ all ${\cal V}^{(j)}\Big(\{{\bf r}_N\}\Big)$ are non-negative, one finds that 
\be
{\cal M}_{\{{\bf S}\}}\Big(\{{\bf
r}_N\}\Big) \leq {\cal S}\Big(\{{\bf
r}_N\}\Big),
\ee
which implies that, quite trivially,  $P_N$ is bounded from below by
\be
P_N \geq E_0\Big\{\exp\Big[ - \; c \; {\cal S}\Big(\{{\bf
r}_N\}\Big) \Big] \Big\}
\ee
On the other hand, for any $p \leq 1$ and any $j > 0$, the difference $1 - p^j \geq 1- p$.
Consequently,  the survival probability is always bounded from above by
\be
P_N \leq  E_0\Big\{\exp\Big[ - \; c \; (1 - p) \; {\cal S}\Big(\{{\bf
r}_N\}\Big) \Big] \Big\},
\ee
i.e. large-$N$ decay of $P_N$ in the model III proceeds slower than the decay in
the ungated trapping problem (section
II) with the  concentration of traps equal to  $c (1 - p)$. This inequality, however,
 does not seem to be
trivial at the first glance,  
since for the model III the factor $c (1 - p)$ represents only the average value of the active traps
concentration, which does fluctuate in time and may exceed $c (1 - p)$ at certain time moments.

Lastly, taking advantage of the analysis in Ref.\cite{donsker}, we infer from 
two latter inequalities that 
the integral effective reaction rate for the model 
III obeys the following two-sided inequality:
\be
\label{label}
a_d \; (1 - p)^{2/(d+2)} \; \Big(\frac{N}{c}\Big)^{d/(d+2)} \; \leq  \; Q^{(III)}_N \; \leq \; 
a_d  \; \Big(\frac{N}{c}\Big)^{d/(d+2)}
\ee
Note now that both lower and upper
 bounds show the same dependence on $N$, (but have slightly different prefactors),
which insures that in the large-$N$ limit the integral effective reaction rate
$Q^{(III)}_N$ grows in proportion to $N^{d/(d+2)}$ and consequently, the
decay of the survival probability for the model III is described by
the dependence $\ln(P_N) \sim - N^{d/(d+2)}$, i.e. the same
 $N$-dependence as in the ungated case \cite{bal,donsker}.

\section{Model IV: A mobile, gated $A$ particle and randomly placed fixed traps}

Consider lastly the trapping model involving an   $A$ particle bearing a $\em stochastic$ gate 
and  performing lattice random walk
in the presence of randomly placed, immobile ungated traps. 
For such a model the indicator function of the reaction event can be written as
\begin{equation}
\Psi_N=\lim_{\beta\to\infty}\exp\left[-\beta\sum_{n=0}^N\eta_n\;\sum_{k=1}^K{\cal I}\left({\bf
r_n}-{\bf S}^{(k)}\right)\right],
\label{eq:56}
\end{equation}
which function equals unity if an $N$-step trajectory $\{{\bf r}_N\}$ does not visit any site from a given 
array $\{{\bf S}^{(k)}\}$ being in the reactive state, and turns
to zero if any of ${\bf
r}_n$, $n = 0,1, \ldots,N$,  coincides with any of ${\bf S}^{(k)}$ when $\eta_n = 1$.

We turn first to  averaging  the indicator function of the reaction event, 
Eq.(\ref{eq:56}),  over 
the traps' placement on the lattice. 
Rewriting first $\Psi_N$ in Eq.(\ref{eq:56}) in  the factorized form
\be
\label{me}
\Psi_N = \prod_{k=1}^K \lim_{\beta\to\infty}\exp\left[-\beta\sum_{n=0}^N\eta_n\; {\cal I}\left({\bf
r_n}-{\bf S}^{(k)}\right)\right],
\ee
and noticing that since all traps are placed independently of each other,  
all multipliers in Eq.(\ref{me}) appear to be statistically independent, we have that the indicator
function averaged over the
traps' placement reads
\be
\Big<\Psi_N\Big>_{\{{\bf S}^{(k)}\}} = 
\Big(\frac{1}{M}\sum_{{\bf S}}\lim_{\beta\to\infty}\exp\Big[-\beta\sum_{n=0}^N\eta_n{\cal 
I}\Big({\bf r}_n-{\bf S}\Big)\Big]\Big)^K, 
\end{equation}
where the brackets with the subscript $\{{\bf S}^{(k)}\}$ stand for the
 averaging with respect to the positions of the traps. Next, turning to the infinite-space limit,
we find that $\Big<\Psi_N\Big>_{\{{\bf S}^{(k)}\}}$ is given explicitly by
\be
\Big<\Psi_N\Big>_{\{{\bf S}^{(k)}\}} = 
\exp\Big\{-c \; {\cal S}\Big(\{{\bf r}_N\}|\{\eta_n=1\}\Big)\},
\label{eq:57}
\end{equation}
where the functional 
\begin{equation}
{\cal S}\Big(\{{\bf r}_N\}|\{\eta_n=1\}\Big) = 
\sum_{{\bf S}}\Big(1-{\cal I}\Big(\sum_{n=0}^N\eta_n{\cal 
I}\Big({\bf r}_n-{\bf S}\Big)\Big)\Big)
\label{eq:570}
\end{equation}
determines  the number of $\em distinct$ sites visited in the $\eta_n = 1$ state by 
a given $N$-step trajectory $\{{\bf r}_N\}$, or, in other words,  the 
number of distinct sites visited by the 
$A$ particle being in the reactive state within a given realization of its $N$-step random walk. 
Below we will study the temporal evolution of the function in 
Eq.(\ref{eq:57}), averaged over the reactivity fluctuations and trajectories $\{{\bf r}_N\}$, 
using first, the Rosenstock approximation,
 and then, by evaluating rigorous lower and upper bounds.
 
\subsection{Rosenstock approximation. An intermediate time decay.}

Consider now the prediction of the Rosenstock-type approximation
 for the $A$ particle survival probability in the model IV. 
Applying the Jensen inequality, we have then that the particle survival probability in the model IV
is bounded by
\be
P_N \geq \exp\Big[ - \; c \; \overline{E_0\Big\{ {\cal S}\Big(\{{\bf r}_N\}|\{\eta_n=1\}\Big) \}}\Big]
\ee
The average of ${\cal S}\Big(\{{\bf r}_N\}|\{\eta_n=1\}\Big)$  
over the reactivity fluctuations can be performed straightforwardly. First of all, we rewrite
 Eq.(\ref{eq:570}) as
\begin{eqnarray}
{\cal S}\Big(\{{\bf r}_N\}|\{\eta_n=1\}\Big) &=& 
\sum_{{\bf S}}\Big(1-{\cal I}\Big(\sum_{n=0}^N\eta_n{\cal 
I}\Big({\bf r}_n-{\bf S}\Big)\Big)\Big) = \nonumber\\
&=&\sum_{{\bf S}}
\Big(1
- \frac{1}{2 \pi} \int^{2 \pi}_0 dZ \; \prod_{n=0}^{N} \exp\Big[ i Z 
\eta_n {\cal I}\Big({\bf r}_n 
- {\bf S}\Big)    \Big] \Big) 
\end{eqnarray}
Next, averaging the latter equation with respect to the distribution of the
variables $\{\eta_n\}$, we have that 
\begin{eqnarray}
\label{IVR}
&&\overline {{\cal S}\Big(\{{\bf r}_N\}|\{\eta_n=1\}\Big)}
= \sum_{{\bf S}}
\Big(1- \frac{1}{2 \pi} \int^{2 \pi}_0 dZ \; \Big(p + 
(1 - p) \exp\Big[i Z\Big]\Big)^{\sum_{n =0}^N {\cal I}\Big({\bf r}_n 
- {\bf S}\Big)}\Big) \equiv \nonumber\\
&\equiv& 
\sum_{{\bf S}} {\cal M}_{{\bf S}}\Big(\{{\bf r}_N\}\Big),
\end{eqnarray}
where the functional ${\cal M}_{{\bf S}}\Big(\{{\bf r}_N\}\Big)$ has been defined
previously in Eq.(\ref{MR}). 

On comparing the result in Eq.(\ref{IVR}) with Eqs.(\ref{rosen}) and (\ref{ap}), 
we notice that
within the Rosenstock approximation the integral effective rate constant 
for the model IV appears to be exactly the same as the one previously found for the model III and
coincides, as well, with the result obtained for the integral effective rate constant in the model I.
Hence, this approximation predicts that the decay of the survival probability in the model III proceeds
exactly in the same fashion as the decay laws obtained
 for the models I and III. In other words, the Rosenstock approximation
appears to be 
completely insensitive to
 the fact which of the reactive species precisely is mobile and which of them
precisely is being
stochastically gated.

\subsection{Large-$N$ decay of the survival probability.}

We start with the derivation of a rigorous lower bound on the $A$ particle survival probability. 
Here, averaging of the indicator function of the reaction event in Eq.(\ref{eq:56}) with respect 
to the reactive state of the
mobile $A$  particle can be performed as follows:
\begin{eqnarray}
\overline{\Psi_N}&=&\overline{\lim_{\beta\to\infty}\exp\left[-\beta \; \sum_{n=0}^N\eta_n\;\sum_{k=1}^K{\cal 
I}\left({\bf r}_n-{\bf  S}^{(k)}\right)\right]} =\nonumber\\
&=&\prod_{n=0}^N\overline{\lim_{\beta\to\infty}\exp\left[-\beta\;\eta_n\;\sum_{k=1}^K{\cal 
I}\left({\bf r}_n-{\bf  S}^{(k)}\right)\right]} = \nonumber\\
&=&\prod_{n=0}^N\left\{(1-p)\lim_{\beta\to\infty}\exp\left[-\beta\sum_{k=1}^K{\cal 
I}\left({\bf r}_n-{\bf  S}^{(k)}\right)\right]+p\right\}
\label{eq:65}
\end{eqnarray}
Next, noticing that
\begin{equation}
(1-p)\lim_{\beta\to\infty}\exp\left[-\beta\sum_{k=1}^K{\cal 
I}\left({\bf r}_n-{\bf  S}^{(k)}\right)\right]+p=\left\{\begin{array}{ll}
1, \;  \mbox{$\sum_{k=1}^K{\cal 
I}\left({\bf r}_n-{\bf  S}^{(k)}\right)=0$,}\\
p, \;  \mbox{otherwise,}
\end{array}
\right.
\label{eq:66}
\end{equation}
and hence,  that
\begin{equation}
(1-p)\lim_{\beta\to\infty}\exp\left[-\beta\sum_{k=1}^K{\cal 
I}\left({\bf r}_n-{\bf  S}^{(k)}\right)\right]+p= \exp\Big[- \alpha_p \Big(1-{\cal I}\left(\sum_{k=1}^K{\cal 
I}\left({\bf r}_n-{\bf  S}^{(k)}\right)\right)\Big)\Big],
\label{eq:67}
\end{equation}
we find that the averaged indicator function of the reaction event obeys
\begin{equation}
\overline{\Psi_N}=\exp[- \alpha_p \; {\cal N}_{\{{\bf S}^{(k)}\}}\Big(\{{\bf r}_N\}\Big)],
\label{eq:68}
\end{equation}
where 
\be
{\cal N}_{\{{\bf S}^{(k)}\}}\Big(\{{\bf r}_N\}\Big) = \sum_{n=0}^N\left(1-{\cal I}\left(\sum_{k=1}^K{\cal 
I}\left({\bf r}_n-{\bf  S}^{(k)}\right)\right)\right)
\ee
is the "residence time"  of a given $N$-step random walk trajectory on the subset of "distinct",
i.e. non-coinciding
sites from the set 
$\{{\bf S}^{(k)}\}$, $k = 1, \ldots, K$. This means that if any two (three and etc) 
sites from $\{{\bf S}^{(k)}\}$ coincide, i.e. the traps overlap, a visit of $\{{\bf
r}_N\}$ to such a multiply covered site singly contributes to ${\cal N}_{\{{\bf S}^{(k)}\}}\Big(\{{\bf r}_N\}\Big)$. 
A rigorous lower bound on $\overline{\Psi_N}$ follows then from an evident inequality
\be
\label{inequality}
\sum_{n=0}^N\left(1-{\cal I}\left(\sum_{k=1}^K{\cal 
I}\left({\bf r}_n-{\bf  S}^{(k)}\right)\right)\right) \leq \sum_{n=0}^N\sum_{k=1}^K{\bf I}\left({\bf
r}_n-{\bf S}^{(k)}\right),
\ee 
where the rhs determines the unconstrained "residence time"  of the
same $N$-step random walk trajectory on  the set 
 of all sites from $\{{\bf S}^{(k)}\}$. Clearly, the inequality in Eq.(\ref{inequality})
 becomes an equality if  all sites 
$\{{\bf S}^{(k)}\}$ are distinct, e.g. when the traps do obey a hard-core exclusion.
Consequently, in virtue of the inequality in 
Eq.(\ref{inequality}), we have that
\begin{equation}
\overline{\Psi_N}\geq \exp\Big[ - \alpha_p \sum_{n=0}^N\sum_{k=1}^K{\cal I}\left({\bf
r}_n-{\bf S}^{(k)}\right)\Big)\Big],
\label{eq:70}
\end{equation}
where the rhs, as one can readily notice, 
is exactly the indicator function of the reaction event for the model III, 
averaged over the reactivity fluctuations.
This
implies the following inequality
\be
\label{lele}
Q^{(IV)}_N \leq Q^{(III)}_N \leq a_d \;  \Big(\frac{N}{c}\Big)^{d/(d+2)},
\ee
which signifies, in particular,  that similarly to the relation between two $Q_N$s, describing
survival of 
gated and ungated targets (models II
and I), the integral effective reaction rate for the model involving a mobile $\em gated$ $A$ 
particle in the
presence of immobile ungated traps is generally less or equal to the  corresponding rate for the model with
 $ungated$ $A$ 
particle performing random walk in presence of gated traps.

We proceed finally to the derivation of a rigorous upper bound on
 the $A$ particle survival probability in the model IV.
To do this, let us turn back
 to the functional ${\cal S}\Big(\{{\bf r}_N\}|\{\eta_n=1\}\Big)$ in Eq.(\ref{eq:570}) 
and recall that it
determines the number of $\em distinct$ sites visited by a particle appearing
 in the $\em reactive$ state within its $N$-step random
walk $\{{\bf r}_N\}$. Note now that similarly to the situation encountered in the derivation of the
analogous bound in the model II, here we do not have any restriction at 
which visit precisely the particle appeared in the reactive state; that is,  each site ${\bf S}$ can be visited
many times by inactive particle until it re-appears at this site eventually being in the reactive state,
which event only does contribute  to the overall value of 
 the functional ${\cal S}\Big(\{{\bf r}_N\}|\{\eta_n=1\}\Big)$. Hence, to find a lower bound on  
${\cal S}\Big(\{{\bf r}_N\}|\{\eta_n=1\}\Big)$ in Eq.(\ref{eq:570}) we will pursue the strategy employed
already in the Section IV, i.e. we will try to restrict the order
of the reactive visit to the site ${\bf S}$. Here, however, it appears to be
 a bit more delicate problem, since we have to deal with
the realization-dependent functional in Eq.(\ref{eq:570}),  rather than with its expected value. 

To find a lower bound on   ${\cal S}\Big(\{{\bf r}_N\}|\{\eta_n=1\}\Big)$, we thus proceed as follows:
First, we constrain the summation in  Eq.(\ref{eq:570}) supposing that it runs not over all sites of an
$\em infinite$ lattice,
but only over some $\em finite$ subset $\{{\bf S}^{*}\}$. Clearly, since the functional 
$\Big(1-{\cal I}\Big(\sum_{n=0}^N\eta_n{\cal 
I}\Big({\bf r}_n-{\bf S}\Big)\Big)\Big)$ is positive-definite, such an operation will result in a lower
bound on ${\cal S}\Big(\{{\bf r}_N\}|\{\eta_n=1\}\Big)$. Next, we define the subset $\{{\bf S}^{*}\}$; we
stipulate that for a given realization of particle's trajectory 
the subset $\{{\bf S}^{*}\}$ is just a collection of 
such lattice sites ${\bf S}$, on which the particle appeared
for the first time being in the reactive state, i.e. sites which remained "virgin" until the first visit in the
$\eta_n = 1$ state.    

More formally, derivation of such a lower bound on ${\cal S}\Big(\{{\bf r}_N\}|\{\eta_n=1\}\Big)$ can be based on the  
evident inequality between the following two
indicator functions:
\begin{eqnarray}
\label{crepe}
&&{\cal I}_{\bf S}\Big(\{{\bf r}_N\}|\{\eta_n=1\}\Big) = \Big( 1-{\cal I}\Big(\sum_{n=0}^N\eta_n{\cal 
I}\Big({\bf r}_n-{\bf S}\Big)\Big) \Big) \geq \nonumber\\
&& \geq \; \Delta_{\bf S}\Big(\{{\bf r}_N\}|\{\eta_n=1\}\Big) = \sum_{n=0}^N \eta_n \Big\{{\cal I}\Big(\sum_{l=0}^{n-1}{\cal 
I}\Big({\bf r}_l-{\bf S}\Big)\Big) - {\cal I}\Big(\sum_{l=0}^{n}{\cal 
I}\Big({\bf r}_l-{\bf S}\Big)\Big)\Big\},
\end{eqnarray}
where the indicator function on the lhs of Eq.(\ref{crepe}) equals one if the site ${\bf S}$ has been
visited at least once by the particle being in the reactive state within a given realization
$\{{\bf r}_N\}$ of its $N$-step walk, and zero - otherwise; while $\Delta_{\bf S}\Big(\{{\bf r}_N\}|\{\eta_n=1\}\Big)$
equals $1$ if within an $N$-step walk the particle appeared for the first time on the site ${\bf S}$ being
in reactive state not having visited this site before; and equals zero otherwise.

Summing both sides of the inequality in Eq.(\ref{crepe}) over all lattice sites ${\bf S}$, we have,
consequently, that
\begin{eqnarray}
{\cal S}\Big(\{{\bf r}_N\}|\{\eta_n=1\}\Big)&=&\sum_{{\bf S}} {\cal I}_{\bf S}\Big(\{{\bf r}_N\}|\{\eta_n=1\}\Big) \geq
\nonumber\\
&\geq& \sum_{{\bf S}} \sum_{n=0}^N \eta_n \Big\{{\cal I}\Big(\sum_{l=0}^{n-1}{\cal 
I}\Big({\bf r}_l-{\bf S}\Big)\Big) - {\cal I}\Big(\sum_{l=0}^{n}{\cal 
I}\Big({\bf r}_l-{\bf S}\Big)\Big)\Big\} = \nonumber\\
&=& \sum_{n=0}^N \eta_n \sum_{{\bf S}} \Big\{\Big(1 - {\cal I}\Big(\sum_{l=0}^{n}{\cal 
I}\Big({\bf r}_l-{\bf S}\Big)\Big)\Big) - \Big(1 - {\cal I}\Big(\sum_{l=0}^{n-1}{\cal 
I}\Big({\bf r}_l-{\bf S}\Big)\Big)\Big)\Big\},
\end{eqnarray}
which yields, in virtue of the definition in Eq.(\ref{virgin}), the desired lower bound of the form
\be
\label{last}
{\cal S}\Big(\{{\bf r}_N\}|\{\eta_n=1\}\Big) \geq \sum_{n = 0}^N \eta_n \Delta\Big(\{{\bf r}_n\}\Big),
\ee
where the rhs of Eq.(\ref{last}) determines the number of "virgin" sites visited by an $N$-step random walk being in the
reactive state.
Equation (\ref{last}) implies, in turn, 
 that the function $\Big<\Psi_N\Big>_{\{{\bf S}^{(k)}\}}$ in Eq.(\ref{eq:57})
 is bounded from above
by
\be
\Big<\Psi_N\Big>_{\{{\bf S}^{(k)}\}} \leq 
\prod_{n=0}^N \exp\Big[ - c \; \eta_n \; \Delta\Big(\{{\bf r}_n\}\Big)\Big]
\ee
Now, averaging both sides of the 
latter equation over the fluctuations of the reactivity, we find
\begin{eqnarray}
\overline{\Big<\Psi_N\Big>}_{\{{\bf S}^{(k)}\}}
 \leq
\overline{\prod_{n=0}^N\exp\left[- c \; \eta_n \;
 \Delta\Big(\{{\bf r}_n\}\Big)\right]}&=&\prod_{n=0}^N\overline{\exp\left[- c \; \eta_n \;
 \Delta\Big(\{{\bf r}_n\}\Big)\right]} =\nonumber\\
&=&\prod_{n=0}^N\left\{ (1-p) \exp\Big[- c \; \Delta\Big(\{{\bf r}_n\}\Big) \Big]+p\right\}
\label{eq:59}
\end{eqnarray}
Recollecting next that the realization-dependent property $\Delta\Big(\{{\bf r}_n\}\Big)$ assumes only two 
values - $1$ or $0$, and hence, that
\begin{equation}
(1-p)\exp\Big[- c \; \Delta\Big(\{{\bf r}_n\}\Big) \Big]+p=\left\{\begin{array}{ll}
\Big\{(1-p)\exp[- c ]+p\Big\}, \;  \mbox{$\Delta\Big(\{{\bf r}_n\}\Big) = 1$,}\\
1, \;  \mbox{$\Delta\Big(\{{\bf r}_n\}\Big) =0$,}
\end{array}
\right.
\label{eq:0}
\end{equation}
we may rewrite quite formally the bound in Eq.(\ref{eq:59}) as
\begin{eqnarray}
\overline{\Big<\Psi_N\Big>}_{\{{\bf S}^{(k)}\}}  &\leq&  \prod_{n=0}^N\left\{(1-p)\exp\Big[- c \; \Delta\Big(\{{\bf r}_n\}\Big) \Big]+p\right\}
= \nonumber\\
&=&\exp\Big[ - \ln\left(\frac{1}{(1-p) \exp[ - c ] +p}\right) \sum_{n=0}^N \Delta\Big(\{{\bf r}_n\}\Big)\Big]
\label{eq:61}
\end{eqnarray}
Lastly, noticing that $\sum_{n=0}^N \Delta\Big(\{{\bf r}_n\}\Big) = {\cal S}\Big(\{{\bf r}_N\}\Big)$,
we find that the $A$ particle survival
probability obeys
\begin{equation}
P_N = E_0\Big\{\overline{\Big<\Psi_N\Big>}_{\{{\bf S}^{(k)}\}}\}  \leq E_0\Big\{\exp\left[-
\ln\left(\frac{1}{(1-p) \exp[ - c ] +p}\right) \; {\cal S}\Big(\{{\bf r}_N\}\Big) \right]\Big\},
\label{eq:63}
\end{equation}
and hence, by taking into account the lower bound in Eq.(\ref{lele}), we arrive at the 
following two-sided inequlity for the integral effective rate
constant in the model IV: 
\be
\label{ts}
\frac{a_d }{c} \; 
\left(\ln\left(\frac{1}{(1-p) \exp[ - c ] +p}\right)\right)^{2/(d+2)}
N^{d/(d+2)} \leq Q^{(IV)}_N \leq a_d \; \Big(\frac{N}{c}\Big)^{d/(d+2)}
\ee
Note that again, both sides of the inequality in Eq.(\ref{ts}) show the same dependence on the time $N$ and consequently, 
determine exactly the $N$-dependence of the integral effective rate
constant in the model IV. We also remark that in the limit $c \ll 1$ one has that 
$\ln(1/((1-p) \exp[ - c ] +p)) \approx (1-p) c$ and hence, in this limit the lower bound
on the integral effective reaction rate for the model IV, Eq.(\ref{ts}), 
coincides with the lower bound on $Q^{(III)}_N$ in Eq.(\ref{label}).

\subsection{Conclusion}

To conclude, we have studied the time
evolution of the $A$ particle survival probability in four models of
stochastically-gated,  diffusion-limited pseudo-first-order reactions of the form $A + B \to B$. 
We have considered two
different models
of target-like annihilation reactions, where the first one
concerns the survival of a single, immobile $A$  particle (the target) in the
presence of randomly moving gated scavengers $B$ (model I), while the second model focuses on the
fate of a gated immobile $A$ particle in the presence of randomly moving ungated
scavengers $B$ (model II). Two other examples of stochastically gated
pseudo-first-order reactions are furnished here by
the trapping  reactions
between a mobile, ungated $A$ particle and a concentration of randomly placed,
immobile,  gated traps $B$ (model III) and the reverse situation with a mobile gated
$A$ particle and randomly placed, immobile, ungated traps $B$ (model IV). 
In all the models studied  we have supposed that
the mobile species perform symmetric lattice random walks. 
Besides, we have adopted here the
two-state Poisson gating model of Ref.\cite{spouge2}, 
in which each of the gates is supposed to be
in either of two states - an active and a blocked one, 
and to update its state at each
tick of the clock at random, independently of the
 previous history as well as of the  gates imposed
on other particles.  

We have demonstrated  that the model I allows for
 an exact solution and
derived the explicit
 asymptotic decay forms
 for lattices of different spatial
dimensionality. Curiously enough, it appeared that for low dimensional lattices ($d
\leq 2$) for which the lattice random walks are recurrent,
 the long time behavior is independent of the
 presence of stochastic gates (as soon as the gating probability $p < 1$)
and proceeds exactly
in the same fashion as for reactions with 
well-defined, non-fluctuating reaction rates (Section II).
Correction terms do, however, depend on the
 gating probability $p$ and may be important
for reactions in which the species are being blocked most of the time.
Next, we have found that for the model I in higher dimensions the decay is described
 by a purely exponential function of time with the
characteristic time dependent on both the gating probability and on the probability of
the eventual return to the origin.  
Physical explanation of the
predicted behavior has been also provided.

Further on, for the model II the decay pattern 
 has been determined exactly in form of rigorous
lower and upper bounds showing the same dependence on the time $N$ but having slightly
different prefactors. We have demonstated that the decay of the $A$ particle survival
probability in this model is characterized by essentially the same time dependence as that for
the model I, i.e. the integral effective reaction rate follows the behavior of the expected
number of distinct sites visited by an $N$-step random walk but may have a different numerical factor.

Next, for models III and  IV we have presented some
approximate results, based on 
the so-called Rosenstock approximation which may provide a plausible description of
the kinetic behavior at intermediate times, as well as 
 exact results concerning the long-time
evolution of the $A$ particle survival probability.  
We have demonstrated that within the Rosenstock 
approach  no difference exists between the kinetic behavior in the 
models III and IV;  Moreover, we have shown that the decay forms coincide with the exact result obtained for
the model I.
Long-time evolution of the decay functions in the models III and IV has been determined 
in form of rigorous lower and upper bounds characterized by the same dependence on the time
$N$. We have also realized that in the case of stochastically-gated trapping reactions the
long-time decay of the $A$ particle survival probability has essentially the same form 
 as that describing the kinetic behavior of
their ungated counterparts (Section II);  
 the characteristic times might be, however,
renormalized to include the dependence on the reaction probability. 

As an interesting by-product of our analysis, we have also shown that the survival
probability in all four models under study can be interpreted as a 
moment generating
function of some refined characteristics of  random walk trajectories. In particular, we have
demonstrated 
that for the model I the survival probability is the moment generating function for
the number of visits rendered by a concentration of independent random walkers to the
origin. In other models this survival probability appears as the moment generating
function of the number of self-intersections of random walk trajectories, residence
time on a disordered array of marked sites, the number of sites visited exactly a
given number of times and so on. Consequently, our results apply as well to  the asymptotical
behavior of the above mentioned generating functions, which in many cases 
has not been
known yet.

\pagebreak

\begin{Large}
Figure Captions.
\end{Large}

\vspace{0.3in}

Fig.1. A schematic illustration of effective geometrical screening of an active $A$
particle by inactive parts of a complex (polymer) molecule. In the situation depicted
on this figure the $A$ particle, which is attached to a polymer, is completely inaccessible
to the $B$ species and the reaction between them is inhibited due solely to the
geometrical restrictions.

Fig.2. In the situation depicted in this figure the
 $B$s may 
diffuse through the
hole opened in the course of polymers' thermal motion
and hence, may enter into reaction with the chemically active
 $A$ particle.

Fig.3. Pseudo-first-order $A + B \to B$ reaction
involving a single mobile
 (immobile) $A$ particle and
 a concentration of
fixed (mobile) $B$ particles taking place
on a  two-dimensional
 lattice. 
The gate may be imposed on either $A$ or $B$ particles.

Fig.4. Representation of the $B$ particles trajectories in form of "directed polymers". Numbers on the $n$-axis
denote the total number of crossings of different points on this axis by different trajectories.

Fig.5. A realization of $N = 130$-step random walk trajectory $\{{\bf r}_{N}\}$ 
on a two-dimensional square lattice.
The sites visited two times are marked by circles; the sites visited three times - by squares and four
times - by diamonds. For this particular realization $\sum_{S} {\cal M}_{\{{\bf S} \}}(\{{\bf
r}_N\})$ is a fourth-order polynomial 
with respect to the gating probability $p$ of the form
$\sum_{S} {\cal M}_{\{{\bf S} \}}(\{{\bf
r}_N\}) = 113 - 69 p - 28 p^2 - 12 p^3 - 4 p^4$. The coefficients in this polynomial are random,
correlated variables dependent on the particular realization of trajectory $\{{\bf r}_{N}\}$.


\begin{references}

\bibitem{perutz} M.F.Perutz and F.S.Mathews, J. Mol. Biology {\bf 21}, 199 (1966)
%
\bibitem{caceres} M.O.Caceres, C.E.Budde and M.A.Re, Phys. Rev. E {\bf  52}, 3462 (1995)
%
\bibitem{igor} T.P{\'a}lszegi, I.M.Sokolov and H.F.Kaufmann, Macromolecules {\bf 31}, 2521 (1998)
%
\bibitem{agmon} N.Agmon and S.Rabinovich, J. Chem. Phys. {\bf 97}, 7270 (1992)
%
\bibitem{zwanzig} R.Zwanzig, J. Phys. Chem. {\bf 97}, 3587 (1992)
%
\bibitem{wang} J.Wang and P.Wolynes, Chem. Phys. {\bf 180}, 141 (1994)
%
\bibitem{klafter} N.Eizenberg and J.Klafter, Chem. Phys. Lett. {\bf 243}, 9 (1995)
%
\bibitem{spouge} J.L.Spouge, J. Virology {\bf 68}, 1782 (1994)
%
\bibitem{mccammon} J.A.McCammon and S.H.Northrup, Nature {\bf 293}, 316 (1981)
%
\bibitem{smol} M. von Smoluchowski,  Z. Phys. Chem. {\bf 92}, 
129 (1917)
%
\bibitem{szabo1} A.Szabo, D.Shoup, S.H..Northrup, and J.A.McCammon, J. Chem. Phys. {\bf 77}, 4484 (1982)
%
\bibitem{szabo2} H.-X. Zhou and A.Szabo, J. Phys. Chem. {\bf 100}, 2597 (1996)
%
\bibitem{ber1} A.M.Berezhkovskii, D.-Y.Yang, S.H.Lin, Yu.A. Makhnovskii, and S.-Y.Sheu,
 J. Chem. Phys. {\bf 106}, 6985 (1997)
%
\bibitem{spouge2} J.L.Spouge, A.Szabo and G.H.Weiss, Phys. Rev. E {\bf  54}, 2248 (1996)
%
\bibitem{ber2} Yu.A. Makhnovskii, A.M.Berezhkovskii, S.-Y.Sheu, D.-Y.Yang, J.Kuo, and S.H.Lin, J. Chem. Phys. {\bf 108}, 971 (1998)
%
\bibitem{ber3} A.M.Berezhkovskii,  D.-Y.Yang,  S.-Y.Sheu and  S.H.Lin, Phys. Rev. E
{\bf 54}, 4462 (1996)
%
\bibitem{weiss} D. Calef and J.M. Deutch, Annu. Rev. Phys. Chem. {\bf 34}, 493 (1983); 
G.H. Weiss, J. Stat. Phys. {\bf 42}, 3 (1986)
%
\bibitem{gleb} G.Oshanin, M.Moreau and S.F.Burlatsky, Adv. Colloid and Interface Sci. {\bf 49} (1994)
%
\bibitem{blumen1}  A. Blumen, G. Zumofen and    J. Klafter, Phys. Rev. B {\bf 30},
5379 (1984)
%
\bibitem{blumen}  A. Blumen, J. Klafter and G. Zumofen,  
in: Optical Spectroscopy of Glasses, ed.: 
I. Zschokke, (Reidel Publ., Dordrecht, 1986)
%
\bibitem{burlatsky} S.F. Burlatsky and A.A. Ovchinnikov, Sov. Phys. JETP {\bf 65}, 
908 (1987)
%
\bibitem{szabo3} A.Szabo, R.Zwanzig and N.Agmon, Phys. Rev. Lett. {\bf 61}, 2496 (1989)
%
\bibitem{redner} S.Redner and K.Kang, J. Phys. A {\bf 17}, L451 (1984)
%
\bibitem{rosenstock} H.B.Rosenstock, SIAM J. Appl. Math. {\bf 9}, 169 (1961); Phys. Rev.  {\bf 187}, 1166 (1969)
%
\bibitem{hughes} B.D.Hughes, {\it Random Walks and Random Environments}, (Oxford
Science Publishers, Oxford, 1995)
%
\bibitem{duplantier} B.Duplantier, G.F.Lawler, J.F.Le Gall and T.J.Lyons, Bull. Sc.
Math. {\bf 117}, 91 (1993)
%
\bibitem{lawler} G.F.Lawler, {\it Intersections of Random Walks}, (Birkh{\"a}user,
Boston, 1991)
%
\bibitem{nechaev}  S.K. Nechaev, J. Mod. Phys. B {\bf 4}, 1809 (1990); {\it
Statistics of Knots and Entangled Random Walks},
(WSPC,  Singapore, 1996)       
%
\bibitem{bal} B.Ya.Balagurov and V.T.Vaks, Sov. Phys. JETP {\bf 65}, 1939 (1973)
%
\bibitem{donsker} M.D. Donsker and S.R.S. Varadhan, Comm. Pure Appl. Math. {\bf 
28}, 525 (1975)
%
\bibitem{collins} F.Collins and G.Kimball, J. Colloid Sci. {\bf 4}, 425 (1949)
%
\bibitem{gleb2} G.Oshanin and A.Blumen, J. Chem. Phys. {\bf 108}, 1140 (1998)
%
\bibitem{gleb3} S.F.Burlatsky, A.A.Ovchinnikov and G.Oshanin, 
Sov. Phys. JETP {\bf 68}, 1153 (1989)
%

\end{references}
\end{document}